\documentclass[a4paper,11pt]{article}
\pdfoutput=1
\usepackage{jcappub}

\usepackage{hyperref}
\usepackage[dvipsnames]{xcolor}
\usepackage{amsmath}
\usepackage{amssymb}
\usepackage{mathtools}
\usepackage{siunitx}[=v2]
\usepackage{siunitx}
\usepackage{bigints}
\usepackage{graphicx}

\usepackage{feynmp-auto}

\usepackage{cleveref}

\usepackage{float}
\usepackage{subcaption}
\DeclareSIUnit{\parsec}{\text{pc}}

\allowdisplaybreaks

\newcommand{\baz}{\begin{array}{cc}}

\newcommand{\be}{\begin{equation}}
\newcommand{\ee}{\end{equation}}

\newcommand{\OmegaGW}{\Omega_{\mathrm{GW}}}

\let\oldsqrt\sqrt
\def\sqrt{\mathpalette\DHLhksqrt}
\def\DHLhksqrt#1#2{%
\setbox0=\hbox{$#1\oldsqrt{#2\,}$}\dimen0=\ht0
\advance\dimen0-0.2\ht0
\setbox2=\hbox{\vrule height\ht0 depth -\dimen0}%
{\box0\lower0.4pt\box2}}

\newcommand{\beq}{\begin{equation}}
\newcommand{\eeq}{\end{equation}}
\newcommand{\bea}{\begin{equation}\begin{aligned}}
\newcommand{\eea}{\end{aligned}\end{equation}}

\newlength{\wsingfig}
\newlength{\wdblefig}
\newlength{\wquadfig}
\newlength{\wtriplefig}
\newcommand{\Eq}[1]{Eq.~(\ref{#1})}

\newcommand{\Sec}[1]{Sec.~\ref{#1}}

\begin{document}
\title{Vector induced gravitational waves sourced by primordial magnetic fields}

\author[a]{Arko Bhaumik,}
\author[b,c]{Theodoros Papanikolaou,}
\author[d]{Anish Ghoshal}
\affiliation[a]{Physics and Applied Mathematics Unit, Indian Statistical Institute, \\ 203 B.T. Road,
Kolkata 700108, India}

\affiliation[b]{Scuola Superiore Meridionale, Largo San Marcellino 10, 80138 Napoli, Italy}

\affiliation[c]{Istituto Nazionale di Fisica Nucleare (INFN), Sezione di Napoli, \\ Via Cinthia 21, 80126 Napoli, Italy}

\affiliation[d]{Institute of Theoretical Physics, Faculty of Physics, University of Warsaw, \\ ul. Pasteura 5, 02-093 Warsaw, Poland}

\emailAdd{arkobhaumik@gmail.com}
\emailAdd{t.papanikolaou@ssmeridionale.it}
\emailAdd{anish.ghoshal@fuw.edu.pl}

\abstract{In this work, we develop a generic formalism for the study of cosmological tensor perturbations induced at second order by first order vector metric perturbations, dubbing these induced tensor modes $\textit{vector-induced gravitational waves}$ (VIGWs). In the presence of an active source such as primordial magnetic fields (PMFs), the vector perturbations of the metric do not necessarily decay rapidly with Hubble expansion, but may remain significantly large in amplitude depending upon the background equation-of-state (EoS) parameter, $w$, during the post-inflationary epoch. Considering an inflation-inspired PMF power spectrum of the power-law form $P_B(k)\propto k^{n_\mathrm{B}}$ (where $n_{\rm B}$ is the magnetic spectral index), we show that the VIGW signal is enhanced for a stiff post-inflationary EoS, with the maximum enhancement happening for $w=1$ (kination). The VIGW spectrum exhibits a maximum around the scale crossing the cosmological horizon at the end of reheating, $k_\mathrm{reh}$, while its present day peak amplitude scales as $\Omega_{\rm GW}(k_{\rm reh},\eta_0)\propto \Delta N_{\rm reh}\times(H_{\rm inf}/M_{\rm Pl})^{8}$, where $H_{\rm inf}$ is the Hubble parameter at the end of inflation and $\Delta N_{\rm reh}$ is the duration of the post-inflationary era in $e$-folds. For $w=1$ and $n_{\rm B}>-3/2$, one further obtains a nearly $n_{\rm B}$-independent frequency scaling of the GW spectrum of the form $\Omega_{\rm GW}(f,\eta_0)\propto
\left(\frac{f}{f_{\rm reh}}\right)^{-2.8}$ for $f>f_\mathrm{reh}\equiv k_\mathrm{reh}/(2\pi)$. We also explicitly demonstrate that the VIGW spectrum is dominant over that of the first-order magnetically-sourced GWs from the kination-dominated reheating era. Finally, it is highlighted that the VIGW signal may lie well within the detection bands of several next-generation interferometric GW missions at small scales. Indicatively, for $H_{\rm inf} \sim O(10^{7})\:\mathrm{GeV}$ and $O(10^{14})\:\mathrm{GeV}$, and $\Delta N_{\rm reh} \sim 15$ and $10$, the VIGW signal is expected to be detectable by LISA and ET respectively.}

\keywords{gravitational waves/theory, primordial magnetic fields, cosmological perturbation theory}

\begin{flushright}
\end{flushright}
\maketitle
\section{Introduction}
\label{sec:intro}
The first detection of gravitational waves (GWs) from a binary black hole merger by the Laser Interferometer Gravitational-Wave Observatory (LIGO)~\cite{Abbott:2016blz} came with the promise of opening a new window into the early Universe, which has hitherto remained inaccessible by means of classical electromagnetic (EM) astronomy. Observations of the Cosmic Microwave Background (CMB) by the Planck mission \cite{Akrami:2018odb} have provided us with strong evidence that the initial conditions for a successful Hot Big Bang (HBB) were set by inflation~\cite{Brout:1977ix,Starobinsky:1979ty,Guth:1980zm,Sato:1980yn}. Having said this, our knowledge regarding the final stages of the inflationary period is still limited, \emph{i.e.}, how and when exactly the standard radiation dominated (RD) cosmological epoch was reached via the intermediate epoch of cosmic reheating. At this juncture, GW astronomy possibly enables us to probe directly both astrophysical~\cite{Bailes:2021tot,LISA:2022yao} \emph{and} cosmological events, with the latter consisting of several aspects of cosmic inflation~\cite{Rubakov:1982df,Starobinsky:1983zz}, first-order phase transitions~\cite{1986MNRAS.218..629H,Kosowsky:1992vn,Kosowsky:1992rz,Kamionkowski:1993fg,Caprini:2007xq,Huber:2008hg}, formation and decay of topological defects~\cite{Durrer:1998rw,Dunsky:2021tih,Hiramatsu:2010yz} and properties of primordial black holes as a dark matter candidate in general~\cite{LISACosmologyWorkingGroup:2023njw}. These new physics cases have led to designs and construction proposals of numerous next-generation GW detectors, \emph{e.g.}, pulsar timing arrays (PTA) \cite{Lentati:2015qwp,Shannon:2015ect,Arzoumanian:2015liz,Qin:2018yhy}, LISA \cite{Audley:2017drz}, Taiji \cite{Guo:2018npi}, Tianqin \cite{Luo:2015ght}, DECIGO \cite{Seto:2001qf,Yagi:2011wg}, AION/MAGIS \cite{Badurina:2019hst}, and ET~\cite{Maggiore:2019uih}. These detectors would give us a chance to look at physical phenomena across a wide range of frequencies lying between the $\mathrm{nHz}$ and the $\mathrm{kHz}$ regimes. In particular, any detection of a primordial stochastic GW background at these scales would provide us with an insight into the uncharted territories of theoretically well-motivated cosmic epochs prior to Big Bang Nucleosynthesis (BBN), which, till now, have remained largely opaque to EM probes.

In this regard, an important theoretical candidate for a cosmological GW background are the so-called scalar-induced gravitational waves (SIGWs)~\cite{Matarrese:1992rp,Matarrese:1993zf,Matarrese:1997ay,Carbone:2004iv,Ananda:2006af,Baumann:2007zm}, which have constituted an active field of research in recent years~ \cite{Alabidi:2012ex,Espinosa:2018eve,Kohri:2018awv,Cai:2018dig,Bartolo:2018rku,Inomata:2018epa,Yuan:2019udt,Yuan:2019wwo,Domenech:2017ems, Domenech:2019quo,Ota:2020vfn,Cai:2019jah,Cai:2019amo,Bhattacharya:2019bvk}. Essentially, these are second order tensor perturbations of the metric sourced by first order scalar perturbations~\cite{Kodama:1985bj,Mukhanov:1990me,Noh:2004bc,Hwang:2007ni,Ananda:2006af,Baumann:2007zm} (see ~\cite{Domenech:2021ztg} for a recent review). It is important to remark that the amplitude of the SIGW background depends quadratically on the amplitude of the primordial curvature power spectrum $\mathcal{P}_\mathcal{R}(k)$, which implies an enhancement of the SIGW signal at scales where $\mathcal{P}_\mathcal{R}(k)$ is amplified. Particularly, this possibility is intriguing at small scales with comoving wavenumber $k>10^4\:\mathrm{Mpc}^{-1}$, since the amplitude of $\mathcal{P}_\mathcal{R}(k)$ on scales larger than those probed by CMB and large scale structure (LSS) observations currently remain unconstrained. Therefore, this cosmological GW background may be a significant source of information about the very small scales that exited the Hubble horizon close to the end of inflation~\cite{Assadullahi:2009jc,Bugaev:2010bb,Inomata:2018epa}. In a similar spirit, recent studies have also investigated second order GWs induced by first-order tensors and scalar-tensor mixing~\cite{Gorji:2023ziy,Bari:2023rcw,Picard:2023sbz,Yu:2023lmo}.

In principle, the nonlinear structure of the gravitational field equations allows second order tensor perturbations to be induced by terms that are bilinear in all of the first order perturbations, \emph{i.e.} scalars, vectors, and tensors. Our focus in this paper lies on GWs induced at the second order by first order vector metric perturbations. In the standard cosmological picture, source-free vector perturbations are generally ignored as they decay rapidly with the expansion of the Universe~\cite{Mukhanov:1990me}. However, vector perturbations may be significantly enhanced in presence of active source(s), \emph{e.g.} the anisotropic stress of primordial magnetic fields~\cite{Mack:2001gc,Paoletti:2008ck}, interactions of density perturbations at different scales~\cite{Lu:2007cj,Lu:2008ju}, topological defects such as cosmic strings~\cite{Pogosian:2007gi,Thomas:2009bm,Yamauchi:2012bc}, etc. If produced with a sufficiently high amplitude, these vector modes could then possibly give rise to a detectable vector induced gravitational wave (VIGW) background. 

In this work, as an illustrative example, we consider the specific case of primordial magnetic fields (PMFs) giving rise to enhanced vector perturbations, with the latter then inducing a VIGW background. Interestingly enough, several astrophysical observations carried out over the last few decades have indicated the presence of cosmic magnetic fields permeating the galactic and the intergalactic media. The typically observed $\mathcal{O}(\mathrm{\mu G})$ strength of these magnetic fields on galactic scales~\cite{2004NewAR..48..763V,2011ApJ...728...97V,2019A&A...622A..16O} may be explained based on galactic dynamo models \cite{1950ZNatA...5...65B,Widrow:2002ud,2023ARA&A..61..561B}. On the other hand, the $\mathcal{O}(10^{-16}\:\rm{G})$ lower bound in intergalactic voids arising from the measurements of $\gamma$-ray cascades from TeV-blazars~\cite{2010-Neronov.Vovk-Sci,2011ApJ...727L...4D,Tiede:2017aql} is challenging to explain via astrophysical processes alone. This hints at a possibly primordial origin of these large scale cosmic magnetic fields, \emph{i.e.}, at scenarios of primordial magnetogenesis~\cite{2006MNRAS.370..319B,2013MNRAS.429L..60B,Rodriguez-Medrano:2023uen}. For instance, large-scale PMFs could be of inflationary origin~\cite{1988-Turner.Widrow-PRD,1991-Ratra-Apj.Lett,1993-Dolgov-PRD,2007-Martin.Yokoyama-JCAP}, generated during cosmological phase transitions~\cite{1970MNRAS.147..279H,Vachaspati:1991nm,Kahniashvili:2012uj}, sourced by primordial density perturbations~\cite{Ichiki:2006cd,Naoz:2013wla}, driven by amplification mechanisms of primordial black holes (PBHs)~\cite{Safarzadeh:2017mdy,Papanikolaou:2023nkx}, or even originate from nonlinear electrodynamics \cite{Campanelli:2007cg} as well as in various extra-dimensional models \cite{Atmjeet:2013yta}. Despite the wide variety of proposed production mechanisms, the magnetic field spectra generated from most of these scenarios can be parametrized generically in terms of the magnetic field amplitude at a given time and a magnetic spectral index. As a concrete illustrative example for the magnetic sector, we consider in this work a Ratra-type inflationary magnetogenesis model, which involves the explicit breaking of conformal invariance of the Maxwell action during the inflationary period \cite{Ratra:1991bn}. Apart from this specific example chosen for the purpose of demonstration, a major portion of the analysis in this paper holds generically and applies to any PMF power spectrum of the power-law type.

While our general framework is valid for any arbitrary value of the equation-of-state (EoS) parameter of the background fluid, we focus particularly on VIGWs produced during a post-inflationary era characterized by the barotropic parameter $w=1$ (kination), which is stiffer than the EoS parameter during both the RD era ($w=1/3$) and the matter dominated (MD) era ($w=0$), as the magnetically sourced vector metric perturbations are comparatively enhanced during such a stiff epoch (see \Sec{sec:PMF}). Since the vector perturbations are directly responsible for sourcing the second order VIGW background, we consider such a kination-dominated post-inflationary epoch in this work as the most optimistic possibility to work with, for the purpose of concretely demonstrating a detectable VIGW background. Kination is known to be naturally achieved in several cosmological setups, \emph{e.g.}, in the case of quintessential inflation \cite{Peebles:1998qn, Dimopoulos:2001ix, Akrami:2017cir, Bettoni:2021qfs} and in generic non-oscillatory inflationary frameworks \cite{Ellis:2020krl}. In such scenarios, even after the end of inflation, the inflaton is assumed to roll along its potential for a considerably long period of time. During this period, it transfers a fractional part of its energy density towards the production of visible sector particles, thereby reheating the Universe via mechanisms like gravitational particle production~\cite{Ford:1986sy,Chun:2009yu}, instant preheating \cite{Felder:1998vq,Dimopoulos:2017tud}, curvaton reheating \cite{Feng:2002nb,BuenoSanchez:2007jxm}, Ricci reheating \cite{Dimopoulos:2018wfg,Opferkuch:2019zbd,Bettoni:2021zhq,Laverda:2023uqv,Figueroa:2024asq}, etc.\footnote{Other possible mechanisms are reheating by evaporation of primordial black holes \cite{Dalianis:2021dbs,Bhaumik:2022pil,Bhaumik:2022zdd,Ghoshal:2023fno}, warm quintessential inflation \cite{Dimopoulos:2019gpz,Rosa:2019jci}, and large-scale isocurvature perturbations~\cite{Jiang:2018uce}.} In such scenarios, the early Universe experiences a phase of kination \cite{Joyce:1997fc,Gouttenoire:2021jhk}, during which the kinetic energy of the inflaton acts as the main source of energy density that decreases quickly with Hubble expansion of the Universe as $\rho_\phi\propto a^{-6}$ before the commencement of the RD epoch. 

The paper is organized as follows. In Sec. \ref{sec:PMF}, we briefly review the physics of PMFs, alongside that of the vector perturbations of the metric sourced by the magnetic stress-energy tensor. In Sec. \ref{sec:VIGW}, we first develop the formalism of computing the second-order VIGW background for generic first order vector perturbations, and then specialize to the case of PMF-sourced vector modes during a kination-dominated reheating epoch as a concrete example. In Sec. \ref{sec:Predictions}, we present our numerically computed results for the VIGW spectral abundance, and assess prospects of its detection at future GW interferometer missions that aim to probe the super-$\mu$Hz frequency range. In Sec. \ref{sec:discussion}, we summarize our key results and conclude by pointing out some interesting future directions to explore.

\medskip


\section{Primordial magnetic fields} \label{sec:PMF}

We assume a stochastic non-helical primordial magnetic field, which is characterized by the following two-point correlation function in a homogeneous and isotropic background:
\begin{equation}
\langle B_i(\boldsymbol{k})B_j(\boldsymbol{k}^\prime)\rangle = (2\pi)^3P_{ij}P_B(k)\delta(\boldsymbol{k}+\boldsymbol{k}^\prime),
\end{equation}
where $P_B(k)$ is the comoving magnetic power spectrum defined in terms of the present day magnetic energy density per logarithmic $k$-interval as
\begin{equation}
P_B(k)=\dfrac{(2\pi)^3}{k^3}\dfrac{d\rho_B}{d\ln k}\:,
\end{equation}
and $P_{ij}$ is a projection tensor defined as
\beq\label{eq:P_ij_def}
 P_{ij}= \delta_{ij} - k_i k_j/k^2\:.  
\eeq
The magnetic power spectrum $P_B(k)$ can be parametrized quite generically in a model-agnostic way as a power law $P_B(k) = A_Bk^{n_{\rm B}}$, while the projection tensor is idempotent and transverse satisfying the following properties:
\beq
P_{ij}P_{jk} = P_{ik},\quad k_iP_{ij} = 0.
\eeq
We note here that \Eq{eq:P_ij_def} is valid only for a flat Friedmann-Lema\^{i}tre-Robertson-Walker (FLRW) background, where perturbations can be decomposed into plane wave Fourier components. Introducing thus the Fourier transform convention 
\beq
B_i(\boldsymbol{k}) = \int \mathrm{d}^3\boldsymbol{x} e^{i\boldsymbol{k}\cdot\boldsymbol{x}}B_i(\boldsymbol{x}),
\eeq
the (comoving) magnetic stress-energy tensor can be recast as the convolution integral
\beq\label{eq:tauijb}
\tau^{(B)}_{ij}(\boldsymbol{k}) = \frac{1}{(2\pi)^3}\frac{1}{4\pi}\int\mathrm{d}^3\boldsymbol{p}\left[B_i(\boldsymbol{p})B_j(\boldsymbol{k-p})-\frac{1}{2}\delta_{ij}B_l(\boldsymbol{p})B_l(\boldsymbol{k-p})\right]\:,
\eeq
which can further be decomposed into scalar, vector and tensor parts as 
$\tau^{(B)}_{ij}(\boldsymbol{k})=\Pi^{(S)}_{ij} + \Pi^{(V)}_{ij} + \Pi^{(T)}_{ij}$ through the application of suitable projection operators \cite{cosmpert,Mack:2001gc}.

\subsection{Magnetically induced vector perturbations} \label{subsec:pmfvec}

Having introduced the PMF formalism above, we now briefly review the theoretical setup to describe first order vector metric perturbations sourced by a stochastic PMF~\cite{Caprini:2001nb,Mack:2001gc,Kosowsky:2001xp,Caprini:2006jb,Caprini:2009yp}. As the starting point, the fully perturbed flat FLRW metric with scalar, vector, and tensor fluctuations can be expressed as
\begin{equation}
    ds^2=a(\eta)^2\left[-\left(1+2\Phi\right)d\eta^2-2\zeta_id\eta dx^i+\bigg\{\left(1-2\Psi\right)\delta_{ij}+\left(\partial_i\xi_j+\partial_j\xi_i\right)+\dfrac{1}{2}h_{ij}\bigg\}dx^idx^j\right]\:.
\end{equation}
We thus describe the vector metric perturbations $\delta g^{(V)}_{\mu\nu}$ in momentum-space in terms of the two divergence-less three-vectors $\zeta_i$ and $\xi_i$~\cite{Mukhanov:1990me} as
\beq
\delta g^{(V)}_{0i} = -a^2\zeta_i\:,\:\: \delta g^{(V)}_{ij} = a^2(\xi_i\hat{k}_j + \xi_{j}\hat{k}_i). 
\eeq
Note here that the divergence-free nature of $\zeta_i$ and $\xi_i$ guarantees the absence of associated density perturbations. One now can construct a gauge-invariant vector potential as $V_i = \zeta_i + \dot{\xi}_i/k$ which actually quantifies the vector perturbations of the extrinsic curvature. Gauge freedom allows us to choose a time-independent $\xi$, which yields $\delta g^{(V)}_{0i} = -a^2V_i$. One also needs to account for the vector perturbations of the stress-energy tensor, which can be written in terms of a divergence-free vector $\boldsymbol{v}^{(V)}$ that quantifies a perturbation of the four-velocity $\delta u_\mu = (0,\boldsymbol{v}^{(V)}/a)$, whose background expression in the comoving frame reads $u_\mu = (1,0,0,0)$. Accordingly, one can also construct the vorticity three-vector as 
\beq
\Omega_i = u^{(V)}_i - V_i.
\eeq
Regarding now the evolution of the vector potential, it is straightforward to show from the perturbed Einstein's equations that $V_i$ is governed by \cite{Mack:2001gc}
\beq\label{eq:V_i_evolution}
V^\prime_i(\eta,\boldsymbol{k}) + 2\mathcal{H}(\eta)V_i(\eta,\boldsymbol{k}) = - \frac{16\pi G\Pi^{(V)}_i(\boldsymbol{k})}{a(\eta)^2k},
\eeq
where $\Pi^{(V)}_i(\boldsymbol{k})$ is defined as
\beq
\Pi^{(V)}_i(\boldsymbol{k}) = \Pi^{(V)}_{ij}\hat{k}_j = P_{in} \hat{k}_m\tau^{(B)}_{mn},
\eeq
with the vector part of the comoving magnetic stress-energy tensor given by $\Pi^{(V)}_{ij} = (P_{in}\hat{k}_j+P_{jn}\hat{k}_i)\hat{k}_m\tau^{(B)}_{mn}$. The analytic solution to \eqref{eq:V_i_evolution} is simple, and reads as
\beq\label{eq:V_i_solution}
V_i(\eta,\boldsymbol{k}) = - \frac{16\pi G\Pi^{(V)}_i(\boldsymbol{k})\eta}{a(\eta)^2k}\:.
\eeq
It immediately becomes clear that the temporal evolution of the first order vector perturbations depends crucially on the equation of state (EoS) of the background fluid. In particular, we expect to get comparatively enhanced perturbations on small scales that re-enter the horizon during the post-inflationary reheating period, \emph{if} the latter is temporarily dominated by a stiff EoS ($w>1/3$) before the standard radiation-dominated (RD) era. This occurs due to the time-scaling $V_i\sim\eta/a^2$, where $a\sim\eta^{2/(1+3w)}$, making the vector perturbations evolve as $V_i\sim a^{3(w-1)/2}$ with respect to the scale factor. Hence, stiffer EoS parameters lead to slower decay rates of $V_i$ compared to the RD case $V_i\sim a^{-1}$, while a kination epoch with $w=1$ offers the extreme scenario of $V_i\sim a^0$. Such slowly decaying vector perturbations could then significantly induce gravitational waves at second order in cosmological perturbation theory during a stiff reheating epoch, which is the focus of the present study.

\section{Vector induced gravitational waves (VIGWs)} \label{sec:VIGW}

Let us now present in this section the general mathematical formalism for VIGWs, and subsequently focus on the particular case of a PMF-motivated scenario where such VIGWs can be significantly sourced during a stiff reheating epoch in the early Universe. 

\subsection{General framework} \label{subsec:VIGWgen}

In the Newtonian gauge, the equation of motion for the second order tensor perturbations sourced by first order vector perturbations can be derived using \texttt{xPand} \cite{xpand}, and turns out to be
\beq \label{eq:vecac}
\begin{split}
h_{ij}^{''}(\boldsymbol{x},\eta)+2\mathcal{H}(\eta)h_{ij}^{'}(\boldsymbol{x},\eta)-\Delta h_{ij}(\boldsymbol{x},\eta)=\widehat{\mathcal{T}}^{ab}_{ij}S_{ab}(\boldsymbol{x},\eta)\:,
\end{split}
\eeq
where the overall quadratic vector source term is given by
\begin{eqnarray}\label{eq:Source_VV}
    S_{ij}(\boldsymbol{x},\eta)=-V_a\partial_a(\partial_i V_j+\partial_j V_i)+\partial_aV_i\partial_aV_j+\partial_iV_a\partial_jV_a+2V_a\partial_i\partial_jV_a+\dfrac{\Delta V_i\Delta V_j}{6\mathcal{H}^2(1+w)}\:.
\end{eqnarray}
In the expressions above, $\mathcal{H}(\eta)$ is the conformal Hubble parameter, $w$ is the background equation-of-state (EoS) parameter  and $\widehat{\mathcal{T}}_{ij}^{ab}$ is the transverse-traceless (TT) projection operator. The entire second-order action obtained via \texttt{xPand} also contains couplings between first order scalar and vector modes, which we do not focus on in our present study (see Appendix \ref{sec:appc})
~\footnote{In the case of a scale-invariant scalar power spectrum with a small amplitude of the order of $10^{-9}$ as measured by Planck~\cite{Planck:2018vyg}, one expects that these contributions are small compared to the vector-vector contributions present in \Eq{eq:Source_VV}.}. Transforming \Eq{eq:vecac} to $k$-space involves term-wise convolutions whose details can be found in Appendix \ref{sec:appa}, following which the vector-vector type source of the second order tensor perturbations turns out to be 
\begin{eqnarray}\label{eq:Source_VV_k}
    S_{ij}(\boldsymbol{k},\eta) && = \int\dfrac{d^3q}{(2\pi)^{3/2}}\left[V_i(\boldsymbol{q},\eta)V_j(\boldsymbol{k}-\boldsymbol{q},\eta)\left(\boldsymbol{q}.\left(\boldsymbol{q}-\boldsymbol{k}\right)+\dfrac{q^2|\boldsymbol{q}-\boldsymbol{k}|^2}{6(1+w)\mathcal{H}^2}\right) \right. \nonumber \\
    &&-q_iq_jV_a(\boldsymbol{q},\eta)V_a(\boldsymbol{k}-\boldsymbol{q},\eta)-k_aV_a(\boldsymbol{q},\eta)\bigg(q_iV_j(\boldsymbol{k}-\boldsymbol{q},\eta)+q_jV_i(\boldsymbol{k}-\boldsymbol{q},\eta)\bigg)\bigg]\:,
\end{eqnarray}
where we have retained only the terms which survive the $k$-space TT projection operation $\widehat{\mathcal{T}}^{ab}_{ij}(\boldsymbol{k})=P^{i}_{a}(\boldsymbol{k})P^{j}_{b}(\boldsymbol{k})-\frac{1}{2}P_{ij}(\boldsymbol{k})P^{ab}(\boldsymbol{k})$, where $P_{ij}(\boldsymbol{k})=\delta_{ij}-k_ik_j/k^2$. We now choose to work in the helicity basis, where the symmetric rank-2 polarization tensor can be defined from the polarization vectors $e_{\mp}^i(\boldsymbol{k})$ as
\begin{equation} \label{eq:piij}
    \Pi_{\pm}^{ij}(\boldsymbol{k})\equiv \dfrac{1}{\sqrt{2}}e_{\mp}^i(\boldsymbol{k})e_{\mp}^j(\boldsymbol{k})\:,
\end{equation}
which can be shown to satisfy the identity $\Pi_{\pm}^{ij}(\boldsymbol{k})\widehat{\mathcal{T}}^{ab}_{ij}(\boldsymbol{k})=\Pi_{\pm}^{ab}(\boldsymbol{k})$. Due to orthonormality, $\Pi_{\pm}^{ij}(\boldsymbol{k})$ can be used to extract the tensor mode function as $h_{\pm}(\boldsymbol{k},\eta)=\Pi_{\pm}^{ij}(\boldsymbol{k})h_{ij}(\boldsymbol{k},\eta)$. Hence, the equation for the tensor mode function becomes
\begin{equation} \label{eq:heqhel}
    h_{\pm}^{''}(\boldsymbol{k},\eta)+2\mathcal{H}(\eta)h_{\pm}^{'}(\boldsymbol{k},\eta)+k^2h_{\pm}(\boldsymbol{k},\eta)=\Pi_{\pm}^{ij}(\boldsymbol{k})S_{ij}(\boldsymbol{k},\eta)= S_{\pm}(\boldsymbol{k},\eta)\:,
\end{equation}
where the helicity-projected source function can be recast as
\begin{eqnarray} \label{eq:slamb}
    &&S_{\lambda}(\boldsymbol{k}_1,\eta_1) \nonumber \\
    &&=\dfrac{1}{\sqrt{2}}\int\dfrac{d^3q_1}{(2\pi)^{3/2}}\left[e_{-\lambda}^\ell(\boldsymbol{k}_1)V_\ell(\boldsymbol{q}_1,\eta_1)e_{-\lambda}^m(\boldsymbol{k}_1)V_m(\boldsymbol{k}_1-\boldsymbol{q}_1,\eta_1)\left(\boldsymbol{q}_1.\left(\boldsymbol{q}_1-\boldsymbol{k}_1\right)+\dfrac{q_1^2|\boldsymbol{q}_1-\boldsymbol{k}_1|^2}{6(1+w)\mathcal{H}_1^2}\right) \right. \nonumber \\
    &&\left.-\left(e_{-\lambda}^\ell(\boldsymbol{k}_1)q_{1\ell}\right)^2V_c(\boldsymbol{q}_1,\eta_1)V_c(\boldsymbol{k}_1-\boldsymbol{q}_1,\eta_1)-2k_{1a}V_a(\boldsymbol{q}_1,\eta_1)e_{-\lambda}^\ell(\boldsymbol{k}_1)q_{1\ell}e_{-\lambda}^m(\boldsymbol{k}_1)V_m(\boldsymbol{k}_1-\boldsymbol{q}_1,\eta_1)\right] \nonumber\:, \\
    &&
\end{eqnarray}
where $\lambda = \pm1$.
The two-point correlation function of the tensor modes is given subsequently by
\begin{eqnarray} \label{eq:h2pt}
    \langle h_{\lambda}(\boldsymbol{k}_1,\eta)h_{\lambda'}(\boldsymbol{k}_2,\eta)\rangle=\int\limits_{\eta_i}^{\eta}d\eta_1\int\limits_{\eta_i}^{\eta}d\eta_2\:g_{k_1}(\eta,\eta_1)g_{k_2}(\eta,\eta_2)\langle S_{\lambda}(\boldsymbol{k}_1,\eta_1)S_{\lambda'}(\boldsymbol{k}_2,\eta_2)\rangle\:,
\end{eqnarray}
where $g_k$ is the Green's function of the tensor modes, being the solution of \Eq{eq:heqhel} with a source term $\delta(\eta-\tilde{\eta})$. For an arbitrary background with a constant EoS parameter $w$, $g_k(\eta,\tilde{\eta})$ reads as 
\begin{equation} \label{eq:greenfunc}
    g_k(\eta,\tilde{\eta})=\dfrac{\pi}{2k}\dfrac{(k\tilde{\eta})^{\alpha+1/2}}{(k\eta)^{\alpha-1/2}}\left[J_{\alpha-1/2}(k\tilde{\eta})Y_{\alpha-1/2}(k\eta)-J_{\alpha-1/2}(k\eta)Y_{\alpha-1/2}(k\tilde{\eta})\right]\:,
\end{equation}
where $J_\alpha(x)$ and $Y_\alpha(x)$ are respectively the Bessel functions of the first and second kind, with $\alpha=2/(1+3w)$ \cite{Domenech_rev}. Thus, one essentially needs to work out the two-point correlator $\langle S_{\lambda}(\boldsymbol{k}_1,\eta_1)S_{\lambda'}(\boldsymbol{k}_2,\eta_2)\rangle$, which lies at the heart of our analysis for the VIGW spectrum. To that end, it is helpful to first expand the vector perturbations in the helicity basis as 
\begin{eqnarray} \label{eq:vimodedec}
    V_i(\boldsymbol{k},\eta)=2\sum\limits_{\lambda=\pm}V_\lambda(\boldsymbol{k},\eta)e_\lambda^i(\boldsymbol{k})\:,
\end{eqnarray}
where the symmetry factor of $2$ arises from the fact that the mode expansion of the vector perturbation in Fourier space contains two additive terms, which can be shown to be identical to each other based on the properties of the polarization vector. One may then use \eqref{eq:vimodedec} to compute the contractions appearing in $S_+(\boldsymbol{k},\eta)$ (See Appendix \ref{sec:appa}). After a lengthy but straightforward calculation, the two-point function of the source term $\langle S_{\lambda}(\boldsymbol{k}_1,\eta_1)S_{\lambda'}(\boldsymbol{k}_2,\eta_2)\rangle$ can be written as a function of contractions of the vector basis  $e_{\mp}^i(\boldsymbol{k})$ and is given by
\begin{eqnarray} \label{eq:S2pt}
    &&\langle S_{\lambda}(\boldsymbol{k}_1,\eta_1)S_{\lambda'}(\boldsymbol{k}_2,\eta_2)\rangle \nonumber \\
    &&= 8\sum\limits_{\lambda_1,..,\lambda_4}\int\dfrac{d^3q_1d^3q_2}{(2\pi)^3}\left[e_{-\lambda}^\ell(\boldsymbol{k}_1)e_{\lambda_1}^\ell(\boldsymbol{q}_1)e_{-\lambda}^m(\boldsymbol{k}_1)e_{\lambda_2}^m(\boldsymbol{k}_1-\boldsymbol{q}_1)\left(\boldsymbol{q}_1.\left(\boldsymbol{q}_1-\boldsymbol{k}_1\right)+\dfrac{q_1^2|\boldsymbol{q}_1-\boldsymbol{k}_1|^2}{6(1+w)\mathcal{H}_1^2}\right) \right. \nonumber \\
    &&\left.-\left(e_{-\lambda}^\ell(\boldsymbol{k}_1)q_{1\ell}\right)^2e_{\lambda_1}^c(\boldsymbol{q}_1)e_{\lambda_2}^c(\boldsymbol{k}_1-\boldsymbol{q}_1)-2e_{-\lambda}^\ell(\boldsymbol{k}_1)q_{1\ell}k_{1c}e_{\lambda_1}^c(\boldsymbol{q}_1)e_{-\lambda}^m(\boldsymbol{k}_1)e_{\lambda_2}^m(\boldsymbol{k}_1-\boldsymbol{q}_1)\right] \nonumber \\
    && \times \left[e_{-\lambda'}^a(\boldsymbol{k}_2)e_{\lambda_3}^a(\boldsymbol{q}_2)e_{-\lambda'}^b(\boldsymbol{k}_2)e_{\lambda_4}^b(\boldsymbol{k}_2-\boldsymbol{q}_2)\left(\boldsymbol{q}_2.\left(\boldsymbol{q}_2-\boldsymbol{k}_2\right)+\dfrac{q_2^2|\boldsymbol{q}_2-\boldsymbol{k}_2|^2}{6(1+w)\mathcal{H}_2^2}\right) \right. \nonumber \\
    &&\left. -\left(e_{-\lambda'}^a(\boldsymbol{k}_2)q_{2a}\right)^2e_{\lambda_3}^d(\boldsymbol{q}_2)e_{\lambda_4}^d(\boldsymbol{k}_2-\boldsymbol{q}_2)-2e_{-\lambda'}^a(\boldsymbol{k}_2)q_{2a}k_{2d}e_{\lambda_3}^d(\boldsymbol{q}_2)e_{-\lambda'}^b(\boldsymbol{k}_2)e_{\lambda_4}^b(\boldsymbol{k}_2-\boldsymbol{q}_2)\right] \nonumber \\
    && \times\langle V_{\lambda_1}(\boldsymbol{q}_1,\eta_1)V_{\lambda_2}(\boldsymbol{k}_1-\boldsymbol{q}_1,\eta_1)V_{\lambda_3}(\boldsymbol{q}_2,\eta_2)V_{\lambda_4}(\boldsymbol{k}_2-\boldsymbol{q}_2,\eta_2)\rangle\:,
\end{eqnarray}
where we have used the notation $\mathcal{H}_1=\mathcal{H}(\eta_1)$ and $\mathcal{H}_2=\mathcal{H}(\eta_2)$.

\subsection{PMF-sourced scenario} \label{subsec:PMFsourced}

Up to this point, our analysis has been valid for generic transverse vector perturbations. Henceforth, we focus on the particular case of a PMF-driven scenario~\cite{Durrer:1999bk,Mack:2001gc,Paoletti:2008ck,2009arXiv0902.1367K}, for which the four-point correlator of the vector mode functions in \eqref{eq:S2pt} can be expanded based on \eqref{eq:V_i_solution}. According to the latter, the vector modes are sourced by the stress-energy tensor of the PMFs as
\begin{eqnarray} \label{eq:vlambdaequation}
    V_\lambda(\boldsymbol{k},\eta)=-\dfrac{16\pi G\eta}{a(\eta)^2k}\:\Pi_\lambda^{(V)}(\boldsymbol{k})\:,
\end{eqnarray}
where $\Pi_\lambda^{(V)}(\boldsymbol{k})= e_{-\lambda}^i(\boldsymbol{k})\Pi_i^{(V)}(\boldsymbol{k})$. As noted in Sec. \ref{subsec:pmfvec}, this implies that $V_\lambda$ becomes constant in time for $w=1$ and decays progressively faster for less stiff values of $w$, which generically makes one expect higher VIGW amplitudes for stiffer EoS-dominated backgrounds. Based on \eqref{eq:vlambdaequation}, the four-point correlator of the vector mode functions is given by
\begin{eqnarray} \label{eq:Vmode4pt}
    &&\langle V_{\lambda_1}(\boldsymbol{q}_1,\eta_1)V_{\lambda_2}(\boldsymbol{k}_1-\boldsymbol{q}_1,\eta_1)V_{\lambda_3}(\boldsymbol{q}_2,\eta_2)V_{\lambda_4}(\boldsymbol{k}_2-\boldsymbol{q}_2,\eta_2)\rangle \nonumber \\
    &&=(16\pi G)^4\dfrac{\eta_1^2\eta_2^2}{a(\eta_1)^4a(\eta_2)^4}\dfrac{1}{q_1q_2|\boldsymbol{k}_1-\boldsymbol{q}_1||\boldsymbol{k}_2-\boldsymbol{q}_2|}e_{-\lambda_1}^{i_1}(\boldsymbol{q}_1)e_{-\lambda_2}^{i_2}(\boldsymbol{k}_1-\boldsymbol{q}_1)e_{-\lambda_3}^{i_3}(\boldsymbol{q}_2)e_{-\lambda_4}^{i_4}(\boldsymbol{k}_2-\boldsymbol{q}_2) \nonumber \\
    &&\times \langle \Pi_{i_1}^{(V)}(\boldsymbol{q}_1)\Pi_{i_2}^{(V)}(\boldsymbol{k}_1-\boldsymbol{q}_1)\Pi_{i_3}^{(V)}(\boldsymbol{q}_2)\Pi_{i_4}^{(V)}(\boldsymbol{k}_2-\boldsymbol{q}_2) \rangle 
\end{eqnarray}
While the exact dependence of the VIGW spectrum on $w$ is hard to obtain analytically due to the complex structures of \eqref{eq:h2pt} and \eqref{eq:greenfunc}, \eqref{eq:Vmode4pt} may give us some important quantitative clues nevertheless. With \eqref{eq:Vmode4pt} in place, the leading time-dependent factor of the integrand appearing in \eqref{eq:h2pt}, \emph{i.e.}, apart from the Bessel functions (whose envelopes are slowly varying) and the Hubble term inside the source function, turns out to be proportional to $\left(\eta_1\eta_2\right)^{\frac{15w-7}{2(1+3w)}}\eta^{\frac{3(w-1)}{1+3w}}$. Hence, as $w$ is lowered from $w=1$ to $w=0$, the magnitude of the integral is expected to be suppressed drastically. The amplitude of the VIGW spectrum should therefore be a monotonically increasing function of $w$, with $w=1$ maximizing the signal strength. 

In order to Wick-expand the four-point correlator in \eqref{eq:Vmode4pt} under the assumption of a Gaussian random field, we recall the following definition of the two-point correlator \cite{Mack:2001gc}:
\begin{eqnarray}
    \langle \Pi_i^{(V)}(\boldsymbol{k})\Pi_j^{(V)}(\boldsymbol{k}')\rangle=P_{ij}(\boldsymbol{k})|\Pi^{(V)}(k)|^2\delta^{(3)}(\boldsymbol{k}+\boldsymbol{k}')\:,
\end{eqnarray}
which finally leads to
\begin{eqnarray} \label{eq:Pi4pt}
    &&\langle \Pi_{i_1}^{(V)}(\boldsymbol{q}_1)\Pi_{i_2}^{(V)}(\boldsymbol{k}_1-\boldsymbol{q}_1)\Pi_{i_3}^{(V)}(\boldsymbol{q}_2)\Pi_{i_4}^{(V)}(\boldsymbol{k}_2-\boldsymbol{q}_2)\rangle \nonumber \\
    &&=\left[ P_{i_1i_3}(q_1)P_{i_2i_4}(|\boldsymbol{k}_1-\boldsymbol{q}_1|)\delta^{(3)}(\boldsymbol{q}_1+\boldsymbol{q}_2)\delta^{(3)}(\boldsymbol{k}_1-\boldsymbol{q}_1+\boldsymbol{k}_2-\boldsymbol{q}_2) \right. \nonumber \\
    &&\left. +P_{i_1i_4}(q_1)P_{i_2i_3}(|\boldsymbol{k}_1-\boldsymbol{q}_1|)\delta^{(3)}(\boldsymbol{k}_1-\boldsymbol{q}_1+\boldsymbol{q}_2)\delta^{(3)}(\boldsymbol{q}_1+\boldsymbol{k}_2-\boldsymbol{q}_2) \right] \nonumber \\
    &&\times|\Pi^{(V)}(q_1)|^2|\Pi^{(V)}(|\boldsymbol{k}_1-\boldsymbol{q}_1|)|^2\:.
\end{eqnarray}
To simplify these expressions further, we need to use specific coordinate representations of the polarization vectors. Fixing a spherical polar basis where the Cartesian components of the momenta are given by $\boldsymbol{k}=(0,0,k)^T$ and $\boldsymbol{q}=q(\sin\theta\cos\phi,\sin\theta\sin\phi,\cos\theta)^T$, the helicity-labelled polarization vectors can be represented as \cite{Cook:2011hg,Cook:2013xea,Cook:2013oaa}
\begin{equation}
    e_{\lambda}^\ell(\boldsymbol{k}_1)=\dfrac{1}{\sqrt{2}}(-\lambda,-i,0)^T\:,
\end{equation}
\begin{equation}
    e_{\lambda}^\ell(\boldsymbol{q}_1)=\dfrac{1}{\sqrt{2}}(-\lambda\cos\theta\cos\phi+i\sin\phi,-\lambda\cos\theta\cos\phi-i\cos\phi,\lambda\sin\theta)^T\:.
\end{equation}
These lead to the useful identity $|e_{\sigma}^i(\boldsymbol{p})e_{\sigma'}^i(\boldsymbol{p'})|^2=\frac{1}{4}\left(1+\sigma\sigma'\frac{\boldsymbol{p}_1.\boldsymbol{p}_2}{p_1p_2}\right)^2$ \cite{Cook:2011hg}, in addition to the representation of the polarization vectors corresponding to $(\boldsymbol{k}_1-\boldsymbol{q}_1)$ which obey the same functional form as above for $\boldsymbol{q}_1$ but with the redefined angles $\tilde{\phi}=\phi+\pi$ and $\tilde{\theta}=\cos^{-1}\left(\frac{k_1-q_1\cos\theta}{|\boldsymbol{k}_1-\boldsymbol{q}_1|}\right)$ \cite{Cook:2013oaa}.

While they are quite straightforward to obtain using a symbolic computation software, we avoid writing here the full analytic expressions of the various combinations of $\langle S_\lambda S_{\lambda'}\rangle$ correlators that result from the relevant substitutions, as their cumbersome forms do not offer additional physical insight. We include only the schematic expression for $\langle S_+S_+\rangle$ in Appendix \ref{sec:appa} for the interested reader. It is worth mentioning here that direct computation reveals $\langle S_+(\boldsymbol{k_1},\eta_1)S_+(\boldsymbol{k_2},\eta_2)\rangle=\langle S_-(\boldsymbol{k_1},\eta_1)S_-(\boldsymbol{k_2},\eta_2)\rangle$ and $\langle S_+(\boldsymbol{k_1},\eta_1)S_-(\boldsymbol{k_2},\eta_2)\rangle=\langle S_-(\boldsymbol{k_1},\eta_1)S_+(\boldsymbol{k_2},\eta_2)\rangle\neq0$. In \Sec{sec:Predictions}, our final results for the present-day GW spectral energy density $\Omega_{\rm GW}(f,\eta_0)$ are obtained by summing over all of these terms which provide similar order of magnitude contributions.

\subsection{The magnetic source term} \label{subsec:magsource}

In principle, \eqref{eq:S2pt}$-$\eqref{eq:Pi4pt} now provide us with all the necessary information regarding the source, and one may readily proceed to compute the VIGW spectrum using \eqref{eq:h2pt} once the explicit functional form of $|\Pi^{(V)}(k)|^2$ is specified. In order thus to proceed, we recall here the power-law parametrization $P_B(k)=A_Bk^{n_{\rm B}}$ of the comoving magnetic power spectrum from \Sec{sec:PMF}, where $n_{\rm B}$ is the magnetic spectral index determined by the underlying PMF generation mechanism, and $A_B$ is a time-independent constant normalized by the present day PMF amplitude smoothed on some given scale. Corresponding to this power-law power spectrum, the convolution source term can be approximated as \cite{Kosowsky:2001xp}
\beq \label{eq:pivsq}
    |\Pi^{(V)}(k)|^2=\int d^3qP_B(k)P_B(|\boldsymbol{k}-\boldsymbol{q}|)\:\approx\:4\pi A_B^2\left[\dfrac{n_{\rm B} k^{2n_B+3}}{(n_{\rm B}+3)(2n_B+3)}+\dfrac{k_{\rm UV}^{2n_B+3}}{2n_B+3}\right]\:,
\eeq
where $k_{\rm UV}$ is an ultraviolet (UV) cutoff for the momentum integral that we identify with the shortest mode that exited the horizon at the end of inflation, \emph{i.e.}, $k_{\rm UV}\sim a_{\rm inf}H_{\rm inf}=k_{\rm inf}$ (the subscript $\mathrm{inf}$ denotes quantities evaluated at the end of inflation). As an example of a particularly well-motivated model, we focus on a Ratra-type inflationary magnetogenesis scenario \cite{Ratra:1991bn} leading to the following form of the comoving magnetic power spectrum parameters:
\beq\label{eq:PMF_spectrum_specific}
    n_{\rm B}=-2s+3\:;\:\:A_B=\dfrac{16}{\pi}\:\Gamma\left(s-\dfrac{1}{2}\right)^2\left(a_{\rm inf}H_{\rm inf}\right)^4\left(\dfrac{1}{2a_{\rm inf}H_{\rm inf}}\right)^{-2(s-3)}\:,
\eeq
where $s$ is the index of the non-conformal coupling function of the electromagnetic action  \cite{Kobayashi:2019uqs}. The scale factor at the end of inflation can be computed using the redshift scaling of the energy density $\rho\sim a^{-3(1+w)}$ in conjunction with the Friedmann scaling $H^2\sim\rho$. Assuming a finite reheating phase (with EoS parameter $w_{\rm reh}$) that lasted for $\Delta N_{\rm reh}$ $e$-folds from the end of inflation and was succeeded by the standard radiation-dominated (RD) epoch, one obtains
\begin{equation} \label{eq:ainf}
    a_{\rm inf}=\left(\dfrac{H_{0}}{H_{\rm eq}}\right)^{2/3}\left(\dfrac{H_{\rm eq}}{H_{\rm inf}}\right)^{1/2}\exp\left[\dfrac{1}{4}(3w_{\rm reh}-1)\Delta N_{\rm reh}\right]\:,
\end{equation}
where $H_0\sim10^{-42}$ GeV is the present day Hubble's constant, and $H_{\rm eq}\sim1.5\times10^{-37}$ GeV is its value at matter-radiation equality \cite{Tomberg:2021ajh}.


\section{The VIGW signal and observational prospects at future detectors} \label{sec:Predictions}

The GW spectral abundance at a given time can be expressed as~\cite{Maggiore:1999vm}
\begin{equation} \label{eq:gwom}
    \Omega_{\rm GW}(k,\eta)=\dfrac{1}{\rho_c}\dfrac{\rm d \rho_{\rm GW}}{{\rm d ln}k}\:\approx\:\dfrac{1}{12}\left[\dfrac{k}{\mathcal{H}(\eta)}\right]^2\mathcal{P}_h(k,\eta)\:,
\end{equation}
where $\rho_{\rm GW}$ is the GW energy density, $\rho_c(\eta)=3H(\eta)^2/(8\pi G)$ is the critical energy density, and $\mathcal{P}_h(k,\eta)$ is the dimensionless tensor power spectrum defined as 
\beq
\langle h_{ij}(\boldsymbol{k},\eta)h^{ij}(\boldsymbol{k'},\eta)\rangle\equiv \frac{2\pi^2}{k^3} \mathcal{P}_h(k,\eta)\delta^{(3)}(\boldsymbol{k}+\boldsymbol{k'}).
\eeq The final approximation in \eqref{eq:gwom} is valid on deeply subhorizon scales $k\gg\mathcal{H}$, where $\dot{h}_\lambda^2\approx k^2h_\lambda^2$ holds for the oscillatory tensor modes. For our current purpose, this is justified as GW production takes place entirely on subhorizon scales after the end of inflation. 

The VIGW signal sourced by the PMF power spectrum needs to be computed at a reference time $\eta=\eta_*$, and then evolved up to $\eta=\eta_0$ today following subsequent cosmic expansion over $\eta_*<\eta<\eta_0$. We consider VIGW production up to the endpoint of the post-inflationary reheating era, \emph{i.e.}, $\eta_*  = \eta_\mathrm{reh}$, when the stiff fluid dominated background transits to the RD era. The rationale behind this choice is provided by the fact that we expect the induced GW energy density spectrum to scale as $k^3$ in the infrared (IR) limit for modes re-entering the horizon during RD, which is generically valid for sufficiently well-behaved stress-energy source terms that are bilinear in dynamical degrees of freedom \cite{Cai:2019cdl}. As the PMF sector under consideration is a particular example of such a source, we expect a peak in the spectrum corresponding to the scale at the end of reheating $k_{\rm reh}=\eta_{\rm reh}^{-1}$, with the signal falling off as $k^3$ for $k\ll k_{\rm reh}$. We depict this expected behaviour of the VIGW spectrum by extrapolating with the aforementioned scaling for $f<f_{\rm reh}$ in subsequent plots, where we have defined the comoving frequency $f=k/(2\pi a_0)$ as usual, where $a_0$ is normalised to unity, i.e. $a_0 =1$.

Furthermore, the onset of RD brings about nonlinear magnetohydrodynamic (MHD) turbulence by rendering the Alfven crossing time comparable to the Hubble time \cite{Banerjee:2004df,Subramanian:2015lua,Sharma:2019jtb}, which is expected to non-trivially affect the subsequent production mechanism of VIGWs. This falls beyond the analytic scope of the present work, since an accurate study of this phenomenon requires computationally expensive MHD simulation-based approaches \cite{RoperPol:2019wvy,Brandenburg:2021pdv,Brandenburg:2021bfx,RoperPol:2022iel}. Thus, for the purpose of demonstration, we currently restrict ourselves to VIGW production only during the stiff period of reheating when the vector mode decays slower than the $a^{-1}$ decay during RD (see Sec. \ref{subsec:pmfvec}), while acknowledging the possibility that this may result in considerable underestimation of the amplitude of the signal at the present time. This also entails an upper limit of $k_{\rm inf}$ and a lower limit of $k_{\rm reh}$ on our momentum integral. The former is imposed as a necessary UV-cutoff since the expression of the magnetic power spectrum generated on superhorizon scales during inflation in \eqref{eq:PMF_spectrum_specific} is not valid for $k>k_{\rm inf}$, while the latter arises from the fact that modes with $k<k_{\rm reh}$ enter during the RD era when nonlinear MHD effects become important (we have further checked that modes with $k<k_{\rm reh}$ in fact contribute negligibly to the momentum integral). However, as we show in this section, a large detectable signal may still be obtained for suitable choices of parameters in spite of the possibility of aforementioned underestimation, which may arise from integrating our expressions only across reheating and neglecting VIGW generation at later times. This finding is encouraging in itself and should motivate further exploration in this direction.

In line with the discussion above, we parametrize the scale factor as
\beq\label{eq:scale_factor_parametrisation}
a(\eta) = a_{\rm reh}\left(\frac{\eta}{\eta_\mathrm{reh}}\right)^{\frac{2}{1+3w_{\rm reh}}}\;\mathrm{for}\;\:\eta_\mathrm{inf}<\eta<\eta_\mathrm{reh},
\eeq
where $a_{\rm reh}=a_{\rm inf}e^{\Delta N_{\rm reh}}$, with $a_{\rm inf}$ given by \eqref{eq:ainf}. The conformal Hubble parameter during reheating then reads $\mathcal{H}(\eta)=\frac{2}{(1+3w_{\rm reh})\eta}$. The scale at the end of reheating may be written as $k_{\rm reh}=\mathcal{H}_{\rm reh}=a_{\rm reh}H_{\rm reh}$. For that, $a_{\rm reh}$ can be computed using \eqref{eq:ainf} alongside the relation $a_{\rm reh}=a_{\rm inf}e^{\Delta N_{\rm reh}}$, and $H_{\rm reh}$ may be related to $H_{\rm inf}$ as $H_{\rm reh}=H_{\rm inf}\exp\left[-3(1+w_{\rm reh})\Delta N_{\rm reh}/2\right]$ using the Friedmann scaling $H^2\propto\rho\propto a^{-3(1+w)}$. One obtains finally that
\begin{equation} \label{eq:kreh}
    k_{\rm reh}=\left(H_{\rm inf}H_{\rm eq}\right)^{1/2}\left(\dfrac{H_{0}}{H_{\rm eq}}\right)^{2/3}\exp\left[-\dfrac{3}{4}(w_{\rm reh}+1)\Delta N_{\rm reh}\right]\:,
\end{equation}
which may be translated to frequency as $f_{\rm reh}=k_{\rm reh}/(2\pi)$.

Considering the radiation energy density to scale as $\rho_r = \frac{\pi^2}{30}g_{*\mathrm{\rho}}T_\mathrm{r}^4$, where the temperature of the primordial plasma ($T_\mathrm{r}$) evolves as $T_\mathrm{r}\propto g^{-1/3}_{*\mathrm{S}}a^{-1}$, the GW spectral abundance at the present epoch is given by
\beq\label{Omega_GW_RD_0}
\Omega_\mathrm{GW}(k,\eta_0) = 
\Omega^{(0)}_r\frac{g_{*\mathrm{\rho},\mathrm{*}}}{g_{*\mathrm{\rho},0}}
\left(\frac{g_{*\mathrm{S},\mathrm{0}}}{g_{*\mathrm{S},\mathrm{*}}}\right)^{4/3}
\OmegaGW(k,\eta_\mathrm{*}),
\eeq
where $g_{*\mathrm{\rho}}$ and $g_{*\mathrm{S}}$ respectively denote the relativistic degrees of freedom of energy and entropy, and $\Omega^{(0)}_r \sim 10^{-4}$ is the present value of the radiation density parameter. Since reheating is considered to have occurred much earlier than BBN, one can show $\frac{g_{*\mathrm{\rho},\mathrm{*}}}{g_{*\mathrm{\rho},0}}\left(\frac{g_{*\mathrm{S},\mathrm{0}}}{g_{*\mathrm{S},\mathrm{*}}}\right)^{4/3}\sim 0.4$. Thus, the GW spectral profile at present essentially traces that at the end of reheating with an overall $\mathcal{O}(10^{-5})$ suppression factor.

In what follows, we work exclusively with a kination-dominated reheating epoch, \emph{i.e.}, $w_{\rm reh}=1$, motivated by the fact that this leads to sustained non-decaying vector perturbations $V_i\sim a^0$ during this epoch (see Sec. \ref{subsec:pmfvec} and \ref{subsec:PMFsourced}). In fact, such a scenario gives us the maximum possible amplitude for VIGWs generated during reheating, with other choices of $w_{\rm reh}<1$ yielding comparatively suppressed spectra due to temporal decay of the vector modes.

\subsection{Predictions for GW detectors} \label{subsec:predGW}

We are now in a position to proceed towards an explicit numerical computation of the VIGW spectral abundance corresponding to different benchmark values of parameters. For that, we first need to maintain consistency with existing bounds, leveraging the CMB and BBN observations, which actually constrain the radiation energy density in the late universe. The latter can be cast in terms of $\Delta N_{\rm eff}=N_{\rm eff}-N_{\rm eff}^{\rm SM}$, which denotes the number of additional degrees of freedom beyond the Standard Model (SM) that are relativistic at the time of recombination, with $N_{\rm eff}^{\rm SM}=3.0440(2)$~\cite{Akita:2020szl,Froustey:2020mcq,Bennett:2020zkv}. As gravitational waves generated adiabatically mimic the effects of free-streaming dark radiation on the CMB power spectra, one consequently obtains an upper limit on the GW spectral abundance at the present time as follows \cite{Yeh:2022heq}:  
\begin{align}
    h^2\Omega_{\rm GW}(f,\eta_0)\lesssim 5.6\times10^{-6}\;\Delta N_\text{eff} \,. \label{eq:darkrad}
\end{align}
This upper bound is valid for $f\gtrsim2\times10^{-11}$ Hz, \emph{i.e.}, for modes which re-entered the horizon prior to BBN. Using the Planck 2018 + BBN bound of $N_{\rm eff}=2.99\pm 0.17$~\cite{ParticleDataGroup:2022pth}, we obtain $\Delta N_{\rm eff}<0.279$ and thus $h^2\bar \Omega_\text{GW} \lesssim  7.2\times 10^{-7}$ at 95\% C.L. We denote this upper bound with the horizontal gray band at the top in the subsequent plots.


Throughout this section, we focus on particular choices of parameters capable of producing VIGW spectra, which can be detectable by next-generation GW interferometer missions at small-scale frequencies $\gtrsim1\mathrm{mHz}$. This frequency range is of particular interest to us, as the peak frequency (which is expected to be located near $f_{\rm reh}$) of our computed VIGW spectra typically falls in the same range for conservatively chosen benchmark values for $H_{\rm inf}$ and $\Delta N_{\rm reh}$. On the other hand, the peak frequency for a viable signal may shift to $\lesssim10^{-6}$ Hz for significantly longer durations of reheating in conjunction with much lowered values of $H_{\rm inf}$, which we do not consider in the present study. As specific examples of both space-based and terrestrial future GW detectors capable of probing the super-$\mu$Hz frequency range, we choose the following six proposed instruments spanning nearly nine decades in frequency space: the Laser Interferometer Space Antenna (LISA) \cite{lisa_1,lisa_2,lisa_3}, the Einstein Telescope (ET) \cite{et_1,et_2,et_3}, the Deci-hertz Interferometer Gravitational Wave Observatory (DECIGO) and its upgraded design (UDECIGO) \cite{decigo_1,decigo_2,decigo_3}, the Big Bang Observer (BBO) \cite{bbo_1,bbo_2,bbo_3}, and the Cosmic Explorer (CE) \cite{ce_1,ce_2,ce_3}. The projected sensitivity curves for these instruments are shown in our subsequent plots, enabling us to directly assess the detectability of the VIGW signal at these instruments in the near future.

\begin{figure}[H]
    \centering
    \includegraphics[width=0.85\textwidth]{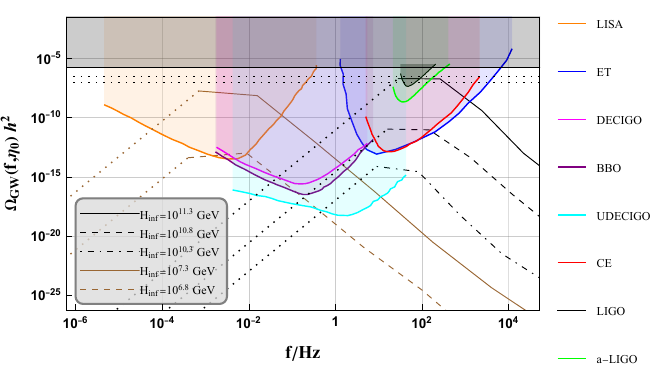}
    \caption{\it Numerically computed profiles of the VIGW spectral abundance at present time $\Omega_{\rm GW}(f,\eta_0)$, corresponding to different benchmark values of $H_{\rm inf}$ for $\Delta N_{\rm reh}=10$ (black) and $\Delta N_{\rm reh}=14$ (brown). For $f<f_{\rm reh}$, the infrared (IR) behaviour has been extrapolated as $(f/f_{\rm reh})^3$, which is the expected IR scaling of $\Omega_{\rm GW}(f,\eta_0)$ for modes re-entering during the radiation-dominated (RD) era. The horizontal gray band on top corresponds to the excluded region based on currently available BBN + $\Delta N_{\rm eff}$ constraints. The dotted horizontal lines just below it respectively denote the projected bounds from the upcoming missions CMB-S4 \cite{Abazajian:2019eic} + CMB-Bharat \cite{Adak:2021lbu} and CMB-HD \cite{CMB-HD:2022bsz}.}
    \label{fig:plotset1}
\end{figure}

\begin{figure}[H]
    \centering
    \includegraphics[width=0.85\textwidth]{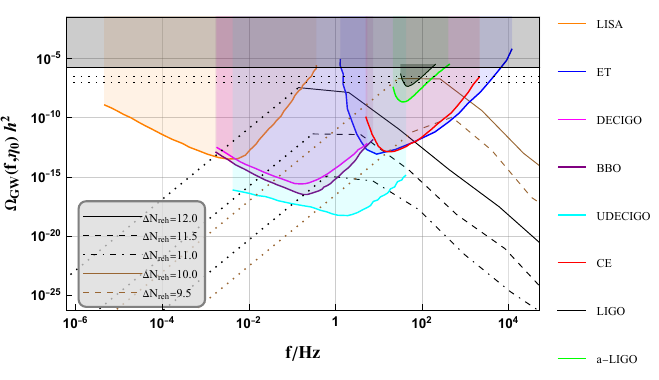}
    \caption{\it Numerically computed profiles of the VIGW spectral abundance at the present time $\Omega_{\rm GW}(f,\eta_0)$, corresponding to different benchmark values of $\Delta N_{\rm reh}$ for $H_{\rm inf}=10^{9.3}$ GeV (black) and $H_{\rm inf}=10^{11.3}$ GeV (brown). Other details are the same as in Fig. \ref{fig:plotset1}.}
    \label{fig:plotset2}
\end{figure}

\begin{figure}[H]
    \centering
    \includegraphics[width=0.85\textwidth]{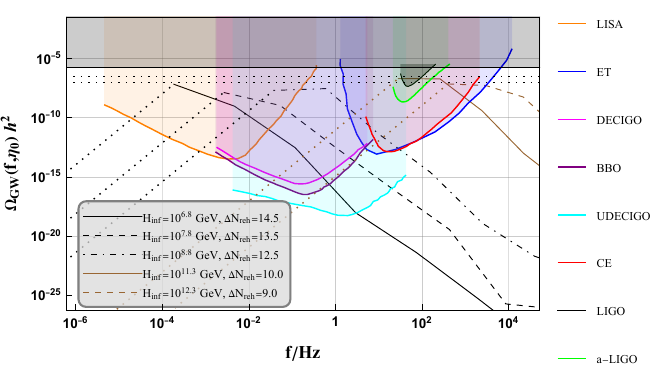}
    \caption{\it Numerically computed profiles of the VIGW spectral abundance at the present time $\Omega_{\rm GW}(f,\eta_0)$, corresponding to different representative combinations of $H_{\rm inf}$ and $\Delta N_{\rm reh}$. The effect of increasing $H_{\rm inf}$ is shown to be antagonistic to that of decreasing $\Delta N_{\rm reh}$, which compensate each other and keep the amplitude nearly unchanged. However, the peak shifts towards higher frequencies, as consistent with increasing $f_{\rm reh}$ for decreasing $\Delta N_{\rm reh}$.}
    \label{fig:plotset3}
\end{figure}

\begin{figure}[H]
    \centering
    \includegraphics[width=0.7\textwidth]{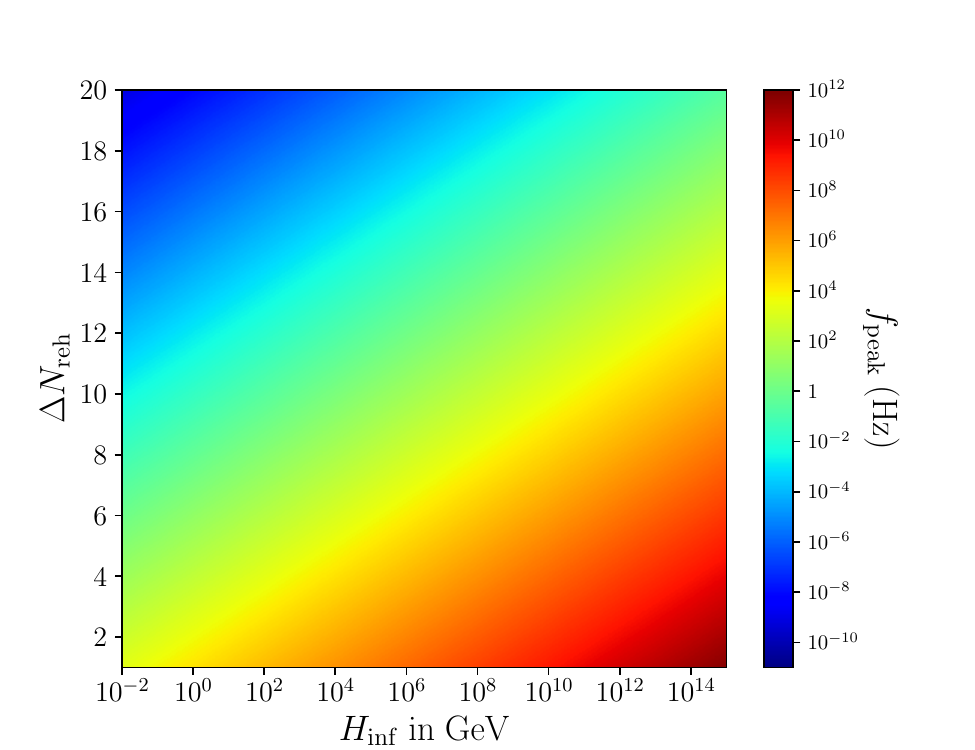}
    \caption{\it Variation of the peak frequency corresponding to the scale at the end of reheating ($f_{\rm peak}$) as a function of the Hubble parameter at the end of inflation ($H_{\rm inf}$) and the duration of reheating ($\Delta N_{\rm reh}$) for a kination-dominated reheating epoch. From an observational point of view, only the range $10^{-6}\:\textrm{Hz}\lesssim f\lesssim10^4\:\textrm{Hz}$ is of interest to us in the light of the next-generation GW interferometer missions.}
    \label{fig:plotfpeak}
\end{figure}

We now move on to our numerically obtained results for $\Omega_{\rm GW}(f,\eta_0)$. We fix the non-conformal coupling parameter at the benchmark value of $s=2$, which yields $n_{\rm B}=-1$ according to \eqref{eq:PMF_spectrum_specific}. In Fig. \ref{fig:plotset1}, we show the variation of $\Omega_{\rm GW}(f,\eta_0)$ with $H_{\rm inf}$, for two different benchmark values of $\Delta N_{\rm reh}=10$ (black curves) and $\Delta N_{\rm reh}=14$ (brown curves). As seen from \eqref{eq:ainf} and \eqref{eq:scale_factor_parametrisation}, $f_{\rm reh}$ is sensitive to both $H_{\rm inf}$ and $\Delta N_{\rm reh}$. Thus, the peak frequency shifts very slightly towards the higher end with slight increase in $H_{\rm inf}$ for a fixed value of $\Delta N_{\rm reh}$, whereas the amplitude increases drastically due to its sharp dependence on $H_{\rm inf}$. On the other hand, increasing $\Delta N_{\rm reh}$ results in a prominent shift of the peak towards the left, which is a direct consequence of having a larger horizon size at the end of reheating. In this case, the values of $H_{\rm inf}$ required for a viable signal need to be lowered as well, so as not to overproduce it beyond the upper bound \eqref{eq:darkrad}. In Fig. \ref{fig:plotset2}, we depict the reverse scenario, by fixing $H_{\rm inf}=10^{9.3}$ GeV (black curves) and $H_{\rm inf}=10^{11.3}$ GeV (brown curves) and focusing on the variation with $\Delta N_{\rm reh}$. In this scenario, increasing $\Delta N_{\rm reh}$ results in gradual leftward shift of the peak in both cases as expected. On the other hand, a higher value of $H_{\rm inf}$ needs to be compensated for by a shorter duration of reheating to have a realistic signal, which implies a smaller horizon size at the end of reheating and thus results in a higher peak frequency. Thereafter, in Fig. \ref{fig:plotset3}, we show the compensatory effect of increasing $H_{\rm inf}$ and decreasing $\Delta N_{\rm reh}$ on the signal amplitude for different pairs of values. While the amplitude remains nearly unchanged, the peak gradually shifts rightward, consistent with the fact that the horizon size at the end of reheating shrinks for shorter reheating epochs, and the change in $H_{\rm inf}$ that may compensate for the amplitude is not sufficient to keep $f_{\rm reh}$ invariant. For the purpose of demonstration, in each of the plots, we have also included a curve which may be ruled out on the basis of the latest observing runs of LIGO \cite{LIGOScientific:2019lzm}. We have also shown the maximum sensitivity threshold for the Advanced LIGO (a-LIGO) design \cite{LIGOScientific:2014pky}, which is already operational. The key message of Figs. \ref{fig:plotset1}$-$\ref{fig:plotfpeak} is that the parameter space under consideration allows us to have a detectable VIGW signal across the \emph{entire} super-$\mu$Hz frequency range targeted by these next-generation GW interferometer missions. 

\begin{figure}[H]
\centering
    \begin{subfigure}[b]{0.48\textwidth}
    \centering
    \includegraphics[width=\textwidth]{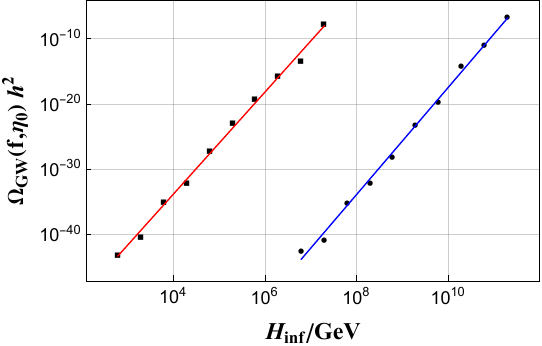}
    \caption{\it $\Delta N_{\rm reh}: 10$ (blue) \& $14$ (red)}
    \end{subfigure}
\quad
    \begin{subfigure}[b]{0.48\textwidth}
    \centering
    \includegraphics[width=\textwidth]{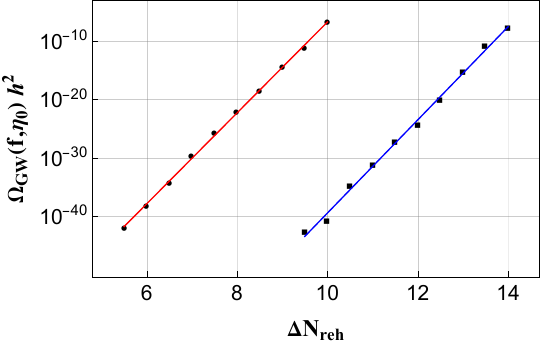}
    \caption{\it $H_{\rm inf}/\textrm{GeV}$: $10^{11.3}$ (blue) \& $10^{7.3}$ (red)}
    \end{subfigure}
\caption{\it Dependence of the amplitude of the present-day VIGW spectral abundance at the peak frequency $\Omega_{\rm GW}(f_{\rm reh},\eta_0)$ on (a) $H_{\rm inf}$ with $\Delta N_{\rm reh}$ held fixed, and (b) $\Delta N_{\rm reh}$ with $H_{\rm inf}$ held fixed. The markers denote the numerically computed values at the corresponding points, which are displayed alongside the best fit straight lines.}
\label{fig:HinfDnrehfits}
\end{figure}

The peak frequency, $f_{\rm peak}\sim f_{\rm reh}$, and the peak amplitude of the signal, $\Omega_{\rm GW}(f_{\rm reh},\eta_0)$, are both found to depend on $H_{\rm inf}$ and $\Delta N_{\rm reh}$ in straightforward manners. In Fig. \ref{fig:plotfpeak}, the dependence of the peak frequency is shown as a function of $H_{\rm inf}$ and $\Delta N_{\rm reh}$, based on the analytic expression of \eqref{eq:kreh}. As seen from the figure, $f_{\rm peak}$ is slightly more sensitive to changes in $\Delta N_{\rm reh}$ compared to those in $H_{\rm inf}$, which also reflects in Figs. \ref{fig:plotset1}$-$\ref{fig:plotset3}. On the other hand, an exact analytic form of the dependence of $\Omega_{\rm GW}(f_{\rm reh},\eta_0)$ on $H_{\rm inf}$ and $\Delta N_{\rm reh}$ is difficult to obtain, as these two parameters enter the functional form of the signal amplitude in complicated ways. However, it is possible to obtain numerical fits by evaluating $\Omega_{\rm GW}(f_{\rm reh},\eta_0)$ at discrete sets of values of $H_{\rm inf}$ and $\Delta N_{\rm reh}$, which we show in Fig. \ref{fig:HinfDnrehfits}. The plots show that the amplitude increases quasi-linearly with both $\log_{10}(H_{\rm inf}/M_{\rm Pl})$ and $\Delta N_{\rm reh}$, and one obtains the approximate scaling relation 
\beq
\Omega_{\rm GW}(f_{\rm reh},\eta_0)\propto \Delta N_{\rm reh}\times(H_{\rm inf}/M_{\rm Pl})^{8}.
\eeq
Since the non-conformal coupling parameter ($s$) directly regulates the magnetic spectral index ($n_{\rm B}$), it is worthwhile asking how sensitive these results are to the choice of $s$, which we have fixed at $s=2$ for our current analysis. In fact, the dependence of the VIGW spectral abundance on $s$ is quite minor, \emph{e.g.}, we find no more than a single order of magnitude suppression in the peak amplitude and no change in the slope of the spectrum if $s=2$ is changed to $s=1$. This may be understood from the form of the source term in \eqref{eq:pivsq}. For $s<9/4$, we have $(2n_B+3)>0$ from \eqref{eq:PMF_spectrum_specific}, which makes the UV-cutoff term dominant in \eqref{eq:pivsq} since $k<k_{\rm UV}$. This suppresses the variation of the spectrum with respect to $n_{\rm B}$ (and hence $s$) as a function of the frequency $f=k/(2\pi)$, since the contribution of the $k^{2n_B+3}$ term itself is negligible. Thus, the only effect is a slight change in the overall amplitude of $\Omega_{\rm GW}(f,\eta_0)$, with no discernible impact on the spectral shape. 

\begin{figure}[H]
    \centering
    \includegraphics[width=0.85\textwidth]{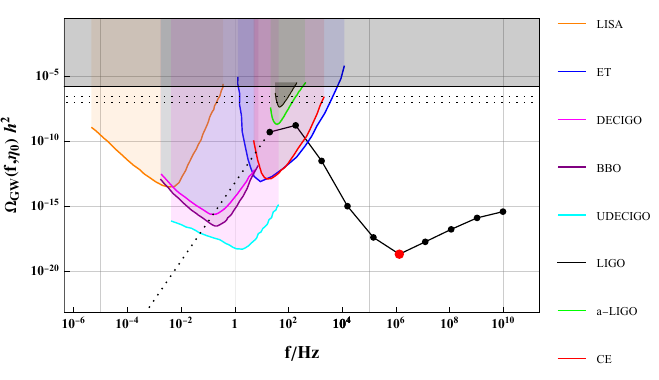}
    \caption{\it The full spectral shape of the VIGW spectral abundance for one representative set of parameter values $H_{\rm inf}=10^{11}$GeV and $\Delta N_{\rm reh}=10$, shown beyond the region of observational interest. The black dots depict the values where the amplitude is numerically computed, and the point highlighted in red marks the critical frequency $f_{\rm crit}$, above which the amplitude rises up to $f_{\rm inf}=k_{\rm inf}/(2\pi)$. This results from the interplay of the different terms in the source spectrum and the $k$-dependent limits of the momentum integral, and is found to be a generic UV feature of all the curves that have been shown across the region of observational interest in the previous plots.}
    \label{fig:fullshape}
\end{figure}

\begin{figure}[H]
\centering
    \begin{subfigure}[b]{0.48\textwidth}
    \centering
    \includegraphics[width=\textwidth]{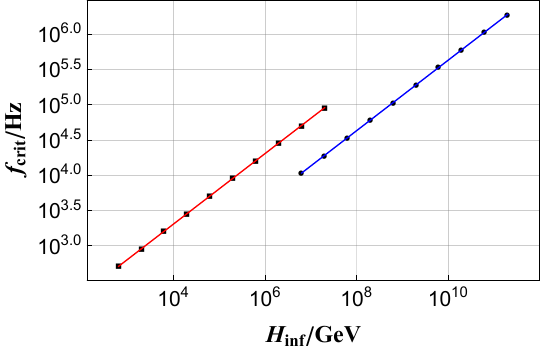}
    \caption{\it $\Delta N_{\rm reh}: 10$ (blue) \& $14$ (red)}
    \end{subfigure}
\quad
    \begin{subfigure}[b]{0.48\textwidth}
    \centering
    \includegraphics[width=\textwidth]{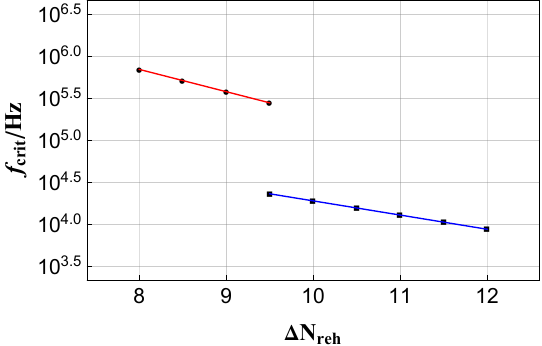}
    \caption{\it $H_{\rm inf}/\textrm{GeV}$: $10^{11.3}$ (blue) \& $10^{7.3}$ (red)}
    \end{subfigure}
\caption{\it Quasi-linear dependence of the critical frequency ($f_{\rm crit}$), above which the amplitude of the VIGW spectral abundance rises monotonically up to $f_{\rm inf}=k_{\rm inf}/(2\pi)$, on $\log_{10}(H_{\rm inf}/M_{\rm Pl})$ and $\Delta N_{\rm reh}$. We find a shift towards higher values as $H_{\rm inf}$ is increased, and the inverse for $\Delta N_{\rm reh}$, with the other parameter kept fixed respectively in each case. The slopes of the best fit straight lines are calculated to be $0.5$ for variation with $\log_{10}(H_{\rm inf}/M_{\rm Pl})$ on the left, and $-0.26$ (red) and $-0.17$ (blue) for variation with $\Delta N_{\rm reh}$ on the right.}
\label{fig:fmindependence}
\end{figure}

Finally, we point out that the particular form of the magnetic source spectrum used in this work, and the interplay of the different momentum-dependent terms therein combined with the $k$-dependent limits of the momentum integral, result in a critical frequency $f_{\rm crit}$, above which the amplitude of the VIGW spectral abundance starts rising monotonically and continues in that manner all the way up to $f_{\rm inf}=k_{\rm inf}/(2\pi)$. This is observed to be a generic UV feature for all the parameter values used for the purpose of illustration in Figs. \ref{fig:plotset1}$-$\ref{fig:plotset3}, where only the region of observational interest has been focused on. On the other hand, the amplitude of the signal at $f_{\rm crit}$ is very small, and the subsequent rise for $f>f_{\rm crit}$ is slow and keeps the signal far below the typical detectability threshold of future experiments. The full spectral shape is shown in Fig. \ref{fig:fullshape}, and the dependence of this turnover frequency on the parameters $H_{\rm inf}$ and $\Delta N_{\rm reh}$ is depicted in Fig. \ref{fig:fmindependence}. We find a quasi-linear dependence on both $\log_{10}(H_{\rm inf}/M_{\rm Pl})$ and $\Delta N_{\rm reh}$ for fiducial parameter values of observational interest. On the other hand, the frequency dependence of the signal across the full range may roughly be parametrized in the following piecewise form, which does not significantly depend on the choice of the fiducial $H_{\rm inf}$ and $\Delta N_{\rm reh}$ values:
\begin{equation}
    \Omega_{\rm GW}(f,\eta_0)\propto
    \begin{cases}
    \left(\frac{f}{f_{\rm reh}}\right)^3\::\:\:0<f<f_{\rm reh}\:, \\
    \left(\frac{f}{f_{\rm reh}}\right)^{-2.8}\::\:\:f_{\rm reh}<f<f_{\rm crit}\:, \\
    \left(\frac{f}{f_{\rm crit}}\right)^{0.8}\::\:\:f_{\rm crit}<f<f_{\rm inf}\:.
    \end{cases}
\end{equation}
It could be interesting to study whether the amplitude at the turning frequency $\Omega_{\rm GW}(f_{\rm crit},\eta_0)$ could somehow be enhanced to observable levels at interferometer missions via secondary mechanisms, or if the UV part of the spectrum at $f>f_{\rm crit}$ could be probed with resonant cavity detectors \cite{Gatti:2024mde}. We do not focus on such prospects in the present study, and defer them to a future work.


\subsection{Comparison with the magnetically sourced first order GW spectrum} \label{subsec:primcomp}

Besides vector perturbations, the magnetic stress-energy tensor is also capable of directly sourcing first order tensor perturbations. Hence, it is worth investigating whether the latter could lead to a competing signal at upcoming detectors. We show here that this is, in fact, not the case, and that the VIGWs indeed provide the dominant contribution to the GW spectral abundance. Following \cite{Mack:2001gc}, the equation for the first order tensor modes generated directly by the magnetic stress-energy tensor can be written as
\begin{equation}
    h_{\lambda}''(\boldsymbol{k},\eta)+2\mathcal{H}(\eta)h_{\lambda}'(\boldsymbol{k},\eta)+k^2h_{\lambda}(\boldsymbol{k},\eta)=\dfrac{8\pi G}{a(\eta)^2}\Pi_\lambda^{ij}(\boldsymbol{k})\Pi_{ij}^{(T)}(\boldsymbol{k})\equiv S_\lambda(\boldsymbol{k},\eta)\:,
\end{equation}
where $\Pi_{ij}^{(T)}(\boldsymbol{k})$ is the tensor projected part of the total magnetic stress-energy tensor, $\Pi_{\lambda}^{ij}(\boldsymbol{k})$ is the helicity-labelled GW polarization tensor, and $\lambda=\pm1$ as earlier. To compute the GW 2-point correlation function, we need $\langle S_{\lambda_1}(\boldsymbol{k}_1,\eta_1)S_{\lambda_2}(\boldsymbol{k_2},\eta_2)\rangle$. This is given by
\begin{equation}
    \langle S_{\lambda_1}(\boldsymbol{k}_1,\eta_1)S_{\lambda_2}(\boldsymbol{k_2},\eta_2)\rangle=\dfrac{(8\pi G)^2}{a(\eta_1)^2a(\eta_2)^2}\Pi_{\lambda_1}^{ij}(\boldsymbol{k_1})\Pi_{\lambda_2}^{ab}(\boldsymbol{k_1})^*\mathcal{M}_{ijab}(k_1)|\Pi^{(T)}(k_1)|^2\delta^{(3)}(\boldsymbol{k_1}+\boldsymbol{k_2})\:,
\end{equation}
where $\mathcal{M}_{ijab}=P_{ia}P_{jb}+P_{ib}P_{ja}-P_{ij}P_{ab}$ with $P_{ij}(k)=\delta_{ij}-k_ik_j/k^2$, and $\Pi^{(T)}(k)$ is the tensor isotropic spectrum for $\Pi_{ij}^{(T)}(\boldsymbol{k})$. Using the properties of the polarization vectors, the contraction in the expression above turns out to be
\begin{equation}
    \Pi_{\lambda_1}^{ij}(\boldsymbol{k_1})\Pi_{\lambda_2}^{ab}(\boldsymbol{k_1})^*\mathcal{M}_{ijab}(k_1)=\delta_{\lambda_1\lambda_2}\:,
\end{equation}
which immediately renders $\langle h_+h_+\rangle=\langle h_-h_-\rangle$ and $\langle h_+h_-\rangle=\langle h_-h_+\rangle=0$. It is then straightforward to finally construct the following dimensionless tensor power spectrum:
\begin{equation} \label{eq:pbtk}
    \mathcal{P}_T^{(B)}(k,\eta)=64G^2k^3|\Pi^{(T)}(k)|^2\left[\int\limits_{\eta_i}^\eta\dfrac{d\tilde\eta}{a(\tilde\eta_1)^2}g_k(\eta,\tilde\eta)\right]^2\:,
\end{equation}
where the Green's function is the same as in \eqref{eq:greenfunc}. 

\begin{figure}[H]
    \centering
    \includegraphics[width=0.85\textwidth]{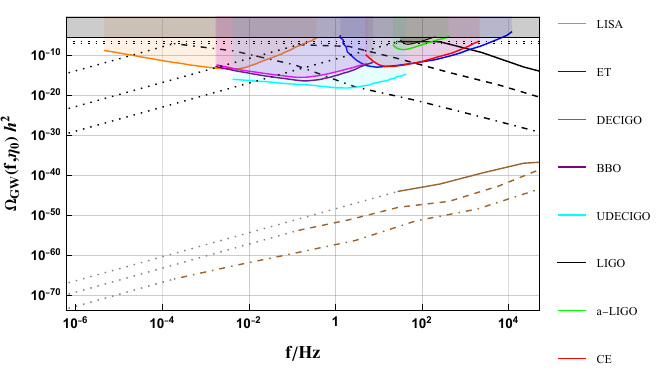}
    \caption{\it Comparison of the GW spectral abundance from the first order tensor modes generated by the magnetic stress-energy tensor (in brown) \textit{vs.} the VIGW spectra (in black) for identical combinations of parameters: $H_{\rm inf}=10^{6.8}$ GeV \& $\Delta N_{\rm reh}=14.5$ (dot-dashed), $H_{\rm inf}=10^{9.3}$ GeV \& $\Delta N_{\rm reh}=12.0$ (dashed), and $H_{\rm inf}=10^{11.3}$ GeV \& $\Delta N_{\rm reh}=10.0$ (solid).}
    \label{fig:plotprimcomp}
\end{figure}

For the purpose of estimating the amplitude of the signal, we can approximate the source term as $|\Pi^{(T)}(k)|^2\approx|\Pi^{(V)}(k)|^2$ following \eqref{eq:pivsq}. Thereafter, to estimate the first order signal complementary to our VIGW spectrum, we evaluate the time integral in \eqref{eq:pbtk} between the end of inflation and the end of reheating for a kination background, and relate the present day GW spectral abundance to it via \eqref{eq:gwom} and \eqref{Omega_GW_RD_0}. The results are shown for a few benchmark combinations of parameter values in Fig. \ref{fig:plotprimcomp}, where the VIGW spectra are observed to be hugely dominant over the first order spectra, with the latter remaining unobservably small corresponding to the same parameter values for which the VIGWs nearly saturate the BBN upper bound.

In fact, it is fairly intuitive to expect the VIGW spectrum to dominate over the first order tensor spectrum based on certain physical considerations. Firstly, the vector modes sourced by the PMF do not decay in amplitude over time throughout the entire duration of VIGW production in a kination-dominated background, as pointed out towards the end of Sec. \ref{subsec:pmfvec}. On the other hand, the energy density of the magnetic field, which directly sources the first order tensor fluctuations, decays as $a^{-4}$ with cosmic expansion in any given background, which may thus lead to a diminished contribution. Secondly, the overall momentum integral in the VIGW power spectrum encapsulates the contribution from all wavelengths between the end of inflation and the end of reheating, at the level of the convolution of the isotropic source spectrum $|\Pi^{(V)}(k)|^2$ with itself. In contrast, the first order tensor power spectrum in \eqref{eq:pbtk} receives no such contribution and remains proportional to $|\Pi^{(T)}(k)|^2$, which seemingly also results in a comparatively lowered amplitude. 

Before concluding this section, we stress again that all the results presented in this section hold specifically for a kination-dominated post-inflationary epoch. Whether the sub-dominant nature of the magnetically sourced first order tensor spectrum is a generic result for arbitrary values of $w_{\rm reh}$ or not lies beyond our current scope. In fact, based on our numerical checks, we found that the amplitude of the first-order spectrum shows a minor dependence on $w_{\rm reh}$, being lowered by $\sim3$ orders of magnitude as $w_{\rm reh}$ decreases from $1$ to $0$. The VIGW spectrum, on the other hand, is expected to have a more pronounced dependence on $w_{\rm reh}$, since the time evolution of the amplitude of the vector perturbations depends crucially on the background. Hence, the VIGW spectral amplitude should decay more sharply in comparison to that of the first order modes, as $w_{\rm reh}$ is lowered from $1$ to $0$ (see discussion in Sec. \ref{subsec:PMFsourced}). Thus, one may reasonably expect a critical value of $w_{\rm reh}$ for which the first order magnetically sourced GW spectrum and the VIGW spectrum could be of comparable strength. Even if that this is the case, Fig. \ref{fig:plotprimcomp} suggests that this would have to take place far below the detectable threshold of the upcoming detectors, unless some post-reheating mechanism during the RD era is capable of significantly boosting the signal strength at a later time. The latter, however, calls for considerably more accurate analyses complemented by post-reheating nonlinear MHD simulation of both the first order signal and the VIGW signal separately, which falls outside the scope of the present work. 

That said, for the kination-dominated reheating epoch under consideration, we note that Fig. \ref{fig:plotprimcomp} indicates a nearly $\sim30$ orders of magnitude hierarchy between the peak amplitudes of the VIGW signal and the first order signal. It is unlikely for any post-reheating process to be capable of bridging this large of a gap and raising the first order signal to a competing level compared to the VIGW one, unless one imagines some very specific mechanism which may suppress the VIGW signal on one hand and enhance the first order signal on the other. On physical grounds, one should not expect MHD turbulence to provide such mutually antagonistic effects on these two signals, whereas further post-RD processes are unlikely to affect the spectral profiles of the freely propagating high-frequency GWs (both first order and VIGW) produced at such early times. To summarize, while MHD turbulence could be a key post-reheating process that may enhance both kinds of signal, we do not expect it to be capable of rendering the first order signal dominant over the VIGW signal for the kination-dominated reheating scenario.

\section{Discussions and Conclusion}
\label{sec:discussion}

In the standard cosmological paradigm, scalar perturbations of the background FLRW metric directly influence the formation of large scale structure in the Universe, whereas source-free tensor perturbations propagate as gravitational waves. In comparison, vector perturbations of the metric are known to decay rapidly in the absence of any active source, which is why their contribution to observable cosmic phenomena is usually taken to be negligible in the minimal scenario. However, non-minimal sources such as primordial magnetic fields may source vector perturbations at a significant level, with the background equation of state playing an important role in the production mechanism and subsequent evolution of the first order vector modes. Among other possible observable features resulting therefrom, these vector perturbations could then be theorized to induce a stochastic gravitational wave background at the second order, in close analogy with the case of scalar induced gravitational waves.

In this study, we have, for the first time, developed the general formalism for studying vector induced gravitational waves at second order in cosmological perturbation theory (Sec. \ref{subsec:VIGWgen}), and applied it to the specific scenario of magnetically generated vector perturbations during a kination-dominated reheating epoch (Sec. \ref{subsec:PMFsourced} and \ref{subsec:magsource}). Considering active VIGW production across this stiff post-inflationary period up to the onset of radiation domination, we demonstrate in Sec. \ref{sec:Predictions} that the VIGW spectral abundance typically peaks around the scale crossing the cosmological horizon at the end of reheating. Due to the nature of our assumptions, we have resorted to a fully numerical computation of the integrals involved in the calculation of the VIGW spectrum, without explicitly providing any intermediate analytic form of the kernel function \emph{per se}. Although it is possible to extract such a kernel by keeping the source spectra aside and evaluating the time integrals analytically for a kination-dominated background, its closed form is found to be too complicated to offer any physical insight. According to our numerical estimates, for modest choices of the Hubble parameter at the end of inflation ($H_{\rm inf}$) and the duration of reheating in $e$-folds ($\Delta N_{\rm reh}$), the peak frequency and the peak amplitude of the signal at the present time fall within the detectability thresholds of several next-generation interferometric GW missions at small scales, \emph{i.e.}, LISA, ET, BBO, DECIGO, UDECIGO, and CE. For instance, for typical benchmark values of $H_{\rm inf} \sim \{O(10^{7}),O(10^{14})\}$ GeV and $\Delta N_{\rm reh} \sim \{{15, 10}\}$, the VIGW signal should be detectable with LISA and ET respectively. We also find an approximate analytical scaling of the GW peak with $H_{\rm inf}$ and $\Delta N_{\rm reh}$ reading $\Omega_{\rm GW}(f_{\rm reh},\eta_0)\propto \Delta N_{\rm reh}\times(H_{\rm inf}/M_{\rm Pl})^{8}$. For $n_B>-3/2$, the sensitivity of the VIGW spectrum to the precise value of $n_B$ is rather suppressed. Furthermore, the VIGW contribution to the GW spectral abundance is found to be dominant over that of the first order tensor modes sourced directly by the magnetic stress-energy tensor, under similar conditions of production and identical choices of parameters (see Sec. \ref{subsec:primcomp}). We have also checked explicitly that the parameter space considered does not result in any backreaction from the PMF sector on the background energy density during reheating (see App. \ref{sec:appb}). This renders the VIGW signal of direct observational relevance at upcoming missions.

At interferometer scales, we also expect the VIGW signal to dominate over the GW background coming from the first order inflationary tensor modes. This is because the first order signal remains constant in amplitude throughout the RD era up to $f_{\rm reh}$, and starts rising as $\sim(f/f_{\rm reh})$ for $f>f_{\rm reh}$ in a kination-dominated background \cite{Figueroa:2019paj,Ghoshal:2024gai}. Thus, it essentially has the lowest amplitude where the VIGW signal has its highest, \emph{i.e.}, at $f=f_{\rm reh}$. For $H_{\rm inf}\sim10^{12}$ GeV, which is about the highest value of $H_{\rm inf}$ we have considered in this work, the amplitude of the inflationary $\Omega_{\rm GW}(f_{\rm reh},\eta_0)$ is around $\mathcal{O}(10^{-19})$ at $f=f_{\rm reh}$, which is completely dominated by the VIGW background. The contribution of the former may only become important at much higher frequencies in the range $f_{\rm reh}<f<f_{\rm crit}$, where the VIGW signal decays as $\sim(f/f_{\rm reh})^{-2.8}$ and the inflationary signal rises as $\sim(f/f_{\rm reh})$.

The present analysis is far from exhaustive, as a primordial magnetic field is only one possible well-motivated candidate for sourcing vector perturbations in the early Universe. On the other hand, the general formalism outlined in Sec. \ref{subsec:VIGWgen} remains valid for any generic first order vector fluctuation of the flat FLRW metric, which may be generated by a variety of other possible sources of primordial origin (see Sec. \ref{sec:intro}), and could possibly also emerge from other scenarios involving vector fields as in the case of vector dark matter (whose induced GW spectrum has recently been studied in Ref. \cite{Marriott-Best:2025sez}). Since the nature of the VIGW spectral abundance inherently depends on the details of the vector source, these other sectors merit investigation, given the encouraging results of the illustrative case that we have considered in this work. Furthermore, while we have focused on VIGW production only during the stiff post-inflationary epoch for the purpose of demonstration, the magnetic production of the vector modes and the subsequent induction of VIGWs need to be studied as well into the radiation dominated era using magnetohydrodynamic (MHD) simulations, in order to arrive at more accurate estimates of the GW signal shape and amplitude, and of the prospect of its detection at future detectors. On the other hand, while post-reheating physical processes such as MHD turbulence might also affect the hierarchy between the VIGW signal strength and the amplitude of the first order magnetically sourced GW spectrum, it is not expected to make the latter dominant over the former for the kination-dominated reheating scenario under consideration (see relevant discussions in Sec. \ref{subsec:primcomp}).

Finally, while we have focused on the pure vector-vector part of the second order source term in this study, the full source for induced GWs also contains self and cross couplings among first order vector, scalar, and tensor modes (see Appendix \ref{sec:appc}). In the general case, all of these terms may not be of negligible magnitude compared to the vector-vector source term, and could contribute to the overall shape of a second order GW spectrum induced by PMF-sourced first order metric perturbations. While evidently beyond the scope of the present work, these other sectors merit detailed investigation, in the light of the encouraging results obtained in this work. We aim thus that the current work will serve as a first demonstration of the potential of first order vector metric perturbations, to induce a second order stochastic gravitational wave background with good detection prospects in the near future, and hence motivate further work in relevant research directions.


\section*{Acknowledgements}

Authors thank Debarun Paul, Supratik Pal, Misao Sasaki, Guillem Domenech and Gianmassimo Tasinato for enlightening discussions. We would also like to thank the anonymous referee, whose comments and feedback have led to substantial improvement of the manuscript. AB thanks CSIR for financial support through Senior Research Fellowship (File no. 09/0093 (13641)/2022-EMR-I). TP and AG acknowledge the contribution of the LISA Cosmology Working Group while TP acknowledges the support of the INFN Sezione di Napoli \textit{iniziativa specifica} QGSKY.

\appendix

\section{Derivation of the two-point function of the vector source} \label{sec:appa}

In the following, we have discarded terms containing any free momentum component ($k_i$ or $k_j$), as only the overall transverse part of $S_{ij}(\boldsymbol{k},\eta)$ survives the TT-projector $\mathcal{T}^{ij}_{ab}(\boldsymbol{k})$. Also, we have exploited the transverse nature of $\boldsymbol{V}$ to simplify, \emph{i.e.} $k_aV_a(\boldsymbol{k},\eta)=0$ has been made use of throughout. Under these conditions, the Fourier transform of \eqref{eq:Source_VV} is given by the sum of the following term-wise convolution integrals (where the subscript ``F'' denotes the Fourier transform):
    \begin{equation}
    \left[-V_a\partial_a(\partial_i V_j+\partial_j V_i)\right]_F=-\int\dfrac{d^3q}{(2\pi)^{3/2}}\left(k_aV_a(\boldsymbol{q})\right)\left[q_i V_j(\boldsymbol{k}-\boldsymbol{q},\eta)+q_j V_i(\boldsymbol{k}-\boldsymbol{q},\eta)\right]\:,
    \end{equation}
    \begin{equation}
    \left[+\partial_aV_i\partial_aV_j\right]_F=\int\dfrac{d^3q}{(2\pi)^{3/2}}\left(q^2-\boldsymbol{k}.\boldsymbol{q}\right)V_i(\boldsymbol{q},\eta)V_j(\boldsymbol{k}-\boldsymbol{q},\eta)\:,
    \end{equation}
    \begin{equation}
    \left[+\partial_iV_a\partial_jV_a\right]_F=\int\dfrac{d^3q}{(2\pi)^{3/2}}q_iq_jV_a(\boldsymbol{q},\eta)V_a(\boldsymbol{k}-\boldsymbol{q},\eta)\:,
    \end{equation}
    \begin{equation}
    \left[+2V_a\partial_i\partial_jV_a\right]_F=-2\int\dfrac{d^3q}{(2\pi)^{3/2}}q_iq_jV_a(\boldsymbol{q},\eta)V_a(\boldsymbol{k}-\boldsymbol{q},\eta)\:,
    \end{equation}
    \begin{equation}
    \left[+\dfrac{\Delta V_i\Delta V_j}{6(1+w)\mathcal{H}^2}\right]_F=\dfrac{1}{6(1+w)\mathcal{H}^2}\int\dfrac{d^3q}{(2\pi)^{3/2}}q^2\left(k^2-2\boldsymbol{k}.\boldsymbol{q}+q^2\right)V_i(\boldsymbol{q},\eta)V_j(\boldsymbol{k}-\boldsymbol{q},\eta)\:.
    \end{equation}
The sum of these five terms yields \eqref{eq:Source_VV_k}. Next, to go from \eqref{eq:Source_VV_k} to \eqref{eq:slamb}, we simply perform the contraction $\Pi^{ij}_\lambda(\boldsymbol{k_1})S_{ij}(\boldsymbol{k_1},\eta_1)$ using the decomposition in \eqref{eq:piij}. Thereafter, using the helicity-basis decomposition of the vector perturbation given in \eqref{eq:vimodedec}, one obtains the following contractions:
\begin{equation}
    e_-^\ell(\boldsymbol{k})V_\ell(\boldsymbol{q},\eta)=2\sum\limits_{\lambda=\pm}e_-^\ell(\boldsymbol{k})e_\lambda^\ell(\boldsymbol{q})V_\lambda(\boldsymbol{q},\eta)\:,
\end{equation}
\begin{equation}
    e_-^m(\boldsymbol{k})V_m(\boldsymbol{k}-\boldsymbol{q},\eta)=2\sum\limits_{\lambda=\pm}e_-^m(\boldsymbol{k})e_\lambda^m(\boldsymbol{k}-\boldsymbol{q})V_\lambda(\boldsymbol{k}-\boldsymbol{q},\eta)\:.
\end{equation}
Using the expressions above, the source function reduces to
\begin{eqnarray} \label{eq:slambexplicitapp}
    &&S_{\lambda}(\boldsymbol{k}_1,\eta_1)=2\sqrt{2}\sum\limits_{\lambda_1,\lambda_2=\pm}\int\dfrac{d^3q_1}{(2\pi)^{3/2}}\bigg[\left(e_{-\lambda}^\ell(\boldsymbol{k}_1)e_{\lambda_1}^\ell(\boldsymbol{q}_1)\right)\left(e_{-\lambda}^m(\boldsymbol{k}_1)e_{\lambda_2}^m(\boldsymbol{k}_1-\boldsymbol{q}_1)\right) \nonumber \\
    &&\times\left(\boldsymbol{q}_1.\left(\boldsymbol{q}_1-\boldsymbol{k}_1\right)+\dfrac{q_1^2|\boldsymbol{q}_1-\boldsymbol{k}_1|^2}{6(1+w)\mathcal{H}_1^2}\right)-\left(e_{-\lambda}^\ell(\boldsymbol{k}_1)q_{1\ell}\right)^2e_{\lambda_1}^a(\boldsymbol{q}_1)e_{\lambda_2}^a(\boldsymbol{k}_1-\boldsymbol{q}_1) \nonumber \\
    &&-2\left(e_{-\lambda}^\ell(\boldsymbol{k}_1)q_{1\ell}\right)\left(k_{1a}e_{\lambda_1}^a(\boldsymbol{q}_1)\right)\left(e_{-\lambda}^m(\boldsymbol{k}_1)e_{\lambda_2}^m(\boldsymbol{k}_1-\boldsymbol{q}_1)\right)\bigg]\times V_{\lambda_1}(\boldsymbol{q}_1,\eta_1)V_{\lambda_2}(\boldsymbol{k}_1-\boldsymbol{q}_1,\eta_1)\:. \nonumber \\
    &&
\end{eqnarray}
It is then straightforward to arrive at \eqref{eq:S2pt} by constructing the unequal time two point correlation of \eqref{eq:slambexplicitapp}. The method for Wick-expanding the vector four-point function in \eqref{eq:S2pt} using the formalism developed in \cite{Mack:2001gc} has been outlined in Sec. \ref{subsec:PMFsourced}. In principle, this is sufficient to obtain the final form of the two-point function of the source. For demonstration, we show below the steps to evaluate the two terms emerging from the Wick expansion of the $\langle S_+S_+\rangle$ correlator, which we denote by $\langle S_+(\boldsymbol{k}_1,\eta_1)S_+(\boldsymbol{k}_2,\eta_2)\rangle_{\rm A}$ and $\langle S_+(\boldsymbol{k}_1,\eta_1)S_+(\boldsymbol{k}_2,\eta_2)\rangle_{\rm B}$. The first one reads
\begin{eqnarray}
        &&\langle S_+(\boldsymbol{k}_1,\eta_1)S_+(\boldsymbol{k}_2,\eta_2)\rangle_A \nonumber \\
        &&= 8(16\pi G)^4\dfrac{\eta_1^2\eta_2^2}{a(\eta_1)^4a(\eta_2)^4}\sum\limits_{\lambda_1,..,\lambda_4}\int\dfrac{d^3q_1}{(2\pi)^{3/2}}\int\dfrac{d^3q_2}{(2\pi)^{3/2}}\dfrac{1}{q_1q_2|\boldsymbol{k}_1-\boldsymbol{q}_1||\boldsymbol{k}_2-\boldsymbol{q}_2|} \nonumber \\
        &&\times\Bigg[\left(\boldsymbol{q}_1.\left(\boldsymbol{q}_1-\boldsymbol{k}_1\right)+\dfrac{q_1^2|\boldsymbol{q}_1-\boldsymbol{k}_1|^2}{6(1+w)\mathcal{H}_1^2}\right)e_{\lambda_1}^\ell(\boldsymbol{q}_1)e_{\lambda_2}^m(\boldsymbol{k}_1-\boldsymbol{q}_1) -q_{1\ell}q_{1m}e_{\lambda_1}^c(\boldsymbol{q}_1)e_{\lambda_2}^c(\boldsymbol{k}_1-\boldsymbol{q}_1) \nonumber \\
        &&-2\left(k_{1c}e_{\lambda_1}^c(\boldsymbol{q}_1)\right)q_{1\ell}e_{\lambda_2}^m(\boldsymbol{k}_1-\boldsymbol{q}_1)\Bigg] \times\Bigg[\left(\boldsymbol{q}_2.\left(\boldsymbol{q}_2-\boldsymbol{k}_2\right)+\dfrac{q_2^2|\boldsymbol{q}_2-\boldsymbol{k}_2|^2}{6(1+w)\mathcal{H}_2^2}\right)e_{\lambda_3}^a(\boldsymbol{q}_2)e_{\lambda_4}^b(\boldsymbol{k}_2-\boldsymbol{q}_2) \nonumber \\
        && -q_{2a}q_{2b}e_{\lambda_3}^d(\boldsymbol{q}_2)e_{\lambda_4}^d(\boldsymbol{k}_2-\boldsymbol{q}_2) -2\left(k_{2d}e_{\lambda_3}^d(\boldsymbol{q}_2)\right)q_{2a}e_{\lambda_4}^b(\boldsymbol{k}_2-\boldsymbol{q}_2)\Bigg]\times \left[e_-^\ell(\boldsymbol{k}_1)e_-^m(\boldsymbol{k}_1)e_-^a(\boldsymbol{k}_2)e_-^b(\boldsymbol{k}_2)\right] \nonumber \\
        &&\times\left[e_{-\lambda_1}^{i_1}(\boldsymbol{q}_1)e_{-\lambda_3}^{i_3}(\boldsymbol{q}_2)P_{i_1i_3}(q_1)\right]\times\left[e_{-\lambda_2}^{i_2}(\boldsymbol{k}_1-\boldsymbol{q}_1)e_{-\lambda_4}^{i_4}(\boldsymbol{k}_2-\boldsymbol{q}_2)P_{i_2i_4}(|\boldsymbol{k}_1-\boldsymbol{q}_1|)\right] \nonumber \\
        &&\times |\Pi^{(V)}(q_1)|^2|\Pi^{(V)}(|\boldsymbol{k}_1-\boldsymbol{q}_1|)|^2\times\delta^{(3)}(\boldsymbol{q}_1+\boldsymbol{q}_2)\times\delta^{(3)}(\boldsymbol{k}_1-\boldsymbol{q}_1+\boldsymbol{k}_2-\boldsymbol{q}_2) \nonumber \\
        && \nonumber \\
        &&= \delta^{(3)}(\boldsymbol{k}_1+\boldsymbol{k}_2)\times\dfrac{8(16\pi G)^4\eta_1^2\eta_2^2}{a(\eta_1)^4a(\eta_2)^4}\sum\limits_{\lambda_1,\lambda_2=\pm1}\int\dfrac{d^3q_1}{(2\pi)^{3/2}}\dfrac{|\Pi^{(V)}(q_1)|^2|\Pi^{(V)}(|\boldsymbol{k}_1-\boldsymbol{q}_1|)|^2}{q_1^2|\boldsymbol{k}_1-\boldsymbol{q}_1|^2} \nonumber \\
        &&\times \left[e_-^\ell(\boldsymbol{k}_1)e_-^m(\boldsymbol{k}_1)e_-^a(\boldsymbol{k}_1)^*e_-^b(\boldsymbol{k}_1)^*\right]\times\Bigg[\left(\boldsymbol{q}_1.\left(\boldsymbol{q}_1-\boldsymbol{k}_1\right)+\dfrac{q_1^2|\boldsymbol{q}_1-\boldsymbol{k}_1|^2}{6(1+w)\mathcal{H}_1^2}\right)e_{\lambda_1}^\ell(\boldsymbol{q}_1)e_{\lambda_2}^m(\boldsymbol{k}_1-\boldsymbol{q}_1) \nonumber \\
        &&-q_{1\ell}q_{1m}e_{\lambda_1}^c(\boldsymbol{q}_1)e_{\lambda_2}^c(\boldsymbol{k}_1-\boldsymbol{q}_1) -2\left(k_{1c}e_{\lambda_1}^c(\boldsymbol{q}_1)\right)q_{1\ell}e_{\lambda_2}^m(\boldsymbol{k}_1-\boldsymbol{q}_1)\Bigg]\times\Bigg[\left(\boldsymbol{q}_1.\left(\boldsymbol{q}_1-\boldsymbol{k}_1\right)+\dfrac{q_1^2|\boldsymbol{q}_1-\boldsymbol{k}_1|^2}{6(1+w)\mathcal{H}_2^2}\right) \nonumber \\
        && \times e_{\lambda_1}^a(\boldsymbol{q}_1)^*e_{\lambda_2}^b(\boldsymbol{k}_1-\boldsymbol{q}_1)^* -q_{1a}q_{1b}e_{\lambda_1}^d(\boldsymbol{q}_1)^*e_{\lambda_2}^d(\boldsymbol{k}_1-\boldsymbol{q}_1)^*-2\left(k_{1d}e_{\lambda_1}^d(\boldsymbol{q}_1)^*\right)q_{1a}e_{\lambda_2}^b(\boldsymbol{k}_1-\boldsymbol{q}_1)^*\Bigg]\:,
    \end{eqnarray}
where, in going to the third step of the equality, we have used $e_{-\lambda_1}^{i_1}(\boldsymbol{q}_1)e_{-\lambda_3}^{i_3}(\boldsymbol{q}_1)^*P_{i_1i_3}(q_1)=\delta_{\lambda_1\lambda_3}$ and $e_{-\lambda_2}^{i_2}(\boldsymbol{k}_1-\boldsymbol{q}_1)e_{-\lambda_4}^{i_4}(\boldsymbol{k}_1-\boldsymbol{q}_1)^*P_{i_2i_4}(|\boldsymbol{k}_1-\boldsymbol{q}_1|)=\delta_{\lambda_2\lambda_4}$. Upon expanding and simplifying, the expression above becomes
    \begin{eqnarray}
        &&\langle S_+(\boldsymbol{k}_1,\eta_1)S_+(\boldsymbol{k}_2,\eta_2)\rangle_A \nonumber \\
        &&= \delta^{(3)}(\boldsymbol{k}_1+\boldsymbol{k}_2)\times\dfrac{8(16\pi G)^4\eta_1^2\eta_2^2}{a(\eta_1)^4a(\eta_2)^4}\sum\limits_{\lambda_1,\lambda_2=\pm1}\int\dfrac{d^3q_1}{(2\pi)^{3/2}}\dfrac{|\Pi^{(V)}(q_1)|^2|\Pi^{(V)}(|\boldsymbol{k}_1-\boldsymbol{q}_1|)|^2}{q_1^2|\boldsymbol{k}_1-\boldsymbol{q}_1|^2} \nonumber \\
        &&\times\Bigg[\left(\boldsymbol{q}_1.\left(\boldsymbol{q}_1-\boldsymbol{k}_1\right)+\dfrac{q_1^2|\boldsymbol{q}_1-\boldsymbol{k}_1|^2}{6(1+w)\mathcal{H}_1^2}\right)\left(\boldsymbol{q}_1.\left(\boldsymbol{q}_1-\boldsymbol{k}_1\right)+\dfrac{q_1^2|\boldsymbol{q}_1-\boldsymbol{k}_1|^2}{6(1+w)\mathcal{H}_2^2}\right) \nonumber \\
        &&\times|e_{\lambda_1}^\ell(\boldsymbol{q}_1)e_-^\ell(\boldsymbol{k}_1)|^2|e_{\lambda_2}^m(\boldsymbol{k}_1-\boldsymbol{q}_1)e_-^m(\boldsymbol{k}_1)|^2+|q_{1\ell}e_-^\ell(\boldsymbol{k}_1)|^4|e_{\lambda_1}^c(\boldsymbol{q}_1)e_{\lambda_2}^c(\boldsymbol{k}_1-\boldsymbol{q}_1)|^2 \nonumber \\
        &&+4|k_{1c}e_{\lambda_1}^c(\boldsymbol{q}_1)|^2|q_{1\ell}e_-^\ell(\boldsymbol{k}_1)|^2|e_-^m(\boldsymbol{k}_1)e_{\lambda_2}^m(\boldsymbol{k}_1-\boldsymbol{q}_1)|^2-2\left(\boldsymbol{q}_1.\left(\boldsymbol{q}_1-\boldsymbol{k}_1\right)+\dfrac{q_1^2|\boldsymbol{q}_1-\boldsymbol{k}_1|^2}{6(1+w)\mathcal{H}_1^2}\right) \nonumber \\
        &&\times\Re\left[\left(q_{1a}e_-^a(\boldsymbol{k}_1)^*\right)^2e_-^\ell(\boldsymbol{k}_1)e_{\lambda_1}^\ell(\boldsymbol{q}_1)e_-^m(\boldsymbol{k}_1)e_{\lambda_2}(\boldsymbol{k}_1-\boldsymbol{q}_1)e_{\lambda_1}^d(\boldsymbol{q})^*e_{\lambda_2}^d(\boldsymbol{k}_1-\boldsymbol{q}_1)^*\right] \nonumber \\
        &&-4\left(\boldsymbol{q}_1.\left(\boldsymbol{q}_1-\boldsymbol{k}_1\right)+\dfrac{q_1^2|\boldsymbol{q}_1-\boldsymbol{k}_1|^2}{6(1+w)\mathcal{H}_1^2}\right)|e_-^m(\boldsymbol{k}_1)e_{\lambda_2}^m(\boldsymbol{k}_1-\boldsymbol{q}_1)|^2\Re\left[k_{1c}e_{\lambda_1}^c(\boldsymbol{q}_1)q_{1\ell}e_-^{\ell}(\boldsymbol{k}_1)e_-^a(\boldsymbol{k}_1)^*e_{\lambda_1}^a(\boldsymbol{q}_1)^*\right] \nonumber \\
        && -4|q_{1\ell}e_-^\ell(\boldsymbol{k}_1)|^2\Re\left[k_{1c}e_{\lambda_1}^c(\boldsymbol{q}_1)e_-^m(\boldsymbol{k}_1)e_{\lambda_2}^m(\boldsymbol{k}_1-\boldsymbol{q}_1)q_{1b}e_-^b(\boldsymbol{k}_1)^*e_{\lambda_1}^d(\boldsymbol{q}_1)^*e_{\lambda_2}^d(\boldsymbol{k}_1-\boldsymbol{q}_1)^*\right] \Bigg]\:,
    \end{eqnarray}
where we have identified $\mathcal{H}_1$ with $\mathcal{H}_2$ in the different additive terms within the integrand, as the entire $\langle S_+S_+\rangle$ correlator later needs to be integrated over $\eta_1$ and $\eta_2$, both from $\eta_0$ to $\eta$. This makes the different additive terms symmetric in $\mathcal{H}_1$ and $\mathcal{H}_2$, but note that the first term contains $\mathcal{H}_1$ and $\mathcal{H}_2$ separately as it is a product term. Some of the contractions appearing in the expression above may be simplified immediately using the representations of the basis vectors given in Sec. \ref{subsec:PMFsourced}, following which $\langle S_+S_+\rangle_A$ takes the form
    \begin{eqnarray}
        &&\langle S_+(\boldsymbol{k}_1,\eta_1)S_+(\boldsymbol{k}_2,\eta_2)\rangle_A \nonumber \\
        &&= \delta^{(3)}(\boldsymbol{k}_1+\boldsymbol{k}_2)\times\dfrac{8(16\pi G)^4\eta_1^2\eta_2^2}{a(\eta_1)^4a(\eta_2)^4}\sum\limits_{\lambda_1,\lambda_2=\pm}\int\dfrac{d^3q_1}{(2\pi)^{3/2}}\dfrac{|\Pi^{(V)}(q_1)|^2|\Pi^{(V)}(|\boldsymbol{k}_1-\boldsymbol{q}_1|)|^2}{q_1^2|\boldsymbol{k}_1-\boldsymbol{q}_1|^2} \nonumber \\
        &&\times\Bigg[\dfrac{1}{16}\left(\boldsymbol{q}_1.\left(\boldsymbol{q}_1-\boldsymbol{k}_1\right)+\dfrac{q_1^2|\boldsymbol{q}_1-\boldsymbol{k}_1|^2}{6(1+w)\mathcal{H}_1^2}\right)\left(\boldsymbol{q}_1.\left(\boldsymbol{q}_1-\boldsymbol{k}_1\right)+\dfrac{q_1^2|\boldsymbol{q}_1-\boldsymbol{k}_1|^2}{6(1+w)\mathcal{H}_2^2}\right) \nonumber \\
        &&\times\left(1+\lambda_1\cos\theta\right)^2\left(1+\lambda_2\dfrac{k_1-q_1\cos\theta}{|\boldsymbol{k}_1-\boldsymbol{q}_1|}\right)^2+\dfrac{1}{16}q_1^4\sin^4\theta\left(1-\lambda_1\lambda_2\dfrac{k_1-q_1\cos\theta}{|\boldsymbol{k}_1-\boldsymbol{q}_1|}\right)^2 \nonumber \\
        &&+\dfrac{1}{4}k_1^2q_1^2\left(1+\lambda_2\dfrac{k_1-q_1\cos\theta}{|\boldsymbol{k}_1-\boldsymbol{q}_1}\right)^2 -2\left(\boldsymbol{q}_1.\left(\boldsymbol{q}_1-\boldsymbol{k}_1\right)+\dfrac{q_1^2|\boldsymbol{q}_1-\boldsymbol{k}_1|^2}{6(1+w)\mathcal{H}_1^2}\right) \nonumber \\
        &&\times \Re\left[\left(q_{1a}e_-^a(\boldsymbol{k}_1)^*\right)^2e_-^\ell(\boldsymbol{k}_1)e_{\lambda_1}^\ell(\boldsymbol{q}_1)e_-^m(\boldsymbol{k}_1)e_{\lambda_2}(\boldsymbol{k}_1-\boldsymbol{q}_1)e_{\lambda_1}^d(\boldsymbol{q})^*e_{\lambda_2}^d(\boldsymbol{k}_1-\boldsymbol{q}_1)^* \right] \nonumber \\
        &&+\dfrac{1}{2}\left(1+\lambda_2\dfrac{k_1-q_1\cos\theta}{|\boldsymbol{k}_1-\boldsymbol{q}_1|}\right)^2\times \Re\left[ k_{1c}e_{\lambda_1}^c(\boldsymbol{q}_1)q_{1\ell}e_-^\ell(\boldsymbol{k}_1)e_-^a(\boldsymbol{k}_1)^*e_{\lambda_1}^a(\boldsymbol{q}_1)^*\right] \nonumber \\
        &&-2q_1^2\sin^2\theta\times\Re\left[k_{1c}e_{\lambda_1}^c(\boldsymbol{q}_1)q_{1b}e_-^b(\boldsymbol{k}_1)^*e_-^m(\boldsymbol{k}_1)e_{\lambda_2}^m(\boldsymbol{k}_1-\boldsymbol{q}_1)e_{\lambda_1}^d(\boldsymbol{q}_1)^*e_{\lambda_2}^d(\boldsymbol{k}_1-\boldsymbol{q}_1)^*\right] \Bigg]\:. \nonumber \\
        &&
    \end{eqnarray}
Similarly, the second piece of the correlator coming from the Wick expansion reads
    \begin{eqnarray}
        &&\langle S_+(\boldsymbol{k}_1,\eta_1)S_+(\boldsymbol{k}_2,\eta_2)\rangle_B \nonumber \\
        &&= 8(16\pi G)^4\dfrac{\eta_1^2\eta_2^2}{a(\eta_1)^4a(\eta_2)^4}\sum\limits_{\lambda_1,..,\lambda_4}\int\dfrac{d^3q_1}{(2\pi)^{3/2}}\int\dfrac{d^3q_2}{(2\pi)^{3/2}}\dfrac{1}{q_1q_2|\boldsymbol{k}_1-\boldsymbol{q}_1||\boldsymbol{k}_2-\boldsymbol{q}_2|} \nonumber \\
        &&\times\Bigg[\left(\boldsymbol{q}_1.\left(\boldsymbol{q}_1-\boldsymbol{k}_1\right)+\dfrac{q_1^2|\boldsymbol{q}_1-\boldsymbol{k}_1|^2}{6(1+w)\mathcal{H}_1^2}\right)e_{\lambda_1}^\ell(\boldsymbol{q}_1)e_{\lambda_2}^m(\boldsymbol{k}_1-\boldsymbol{q}_1) -q_{1\ell}q_{1m}e_{\lambda_1}^c(\boldsymbol{q}_1)e_{\lambda_2}^c(\boldsymbol{k}_1-\boldsymbol{q}_1) \nonumber \\
        &&-2\left(k_{1c}e_{\lambda_1}^c(\boldsymbol{q}_1)\right)q_{1\ell}e_{\lambda_2}^m(\boldsymbol{k}_1-\boldsymbol{q}_1)\Bigg]\times\Bigg[\left(\boldsymbol{q}_2.\left(\boldsymbol{q}_2-\boldsymbol{k}_2\right)+\dfrac{q_2^2|\boldsymbol{q}_2-\boldsymbol{k}_2|^2}{6(1+w)\mathcal{H}_2^2}\right)e_{\lambda_3}^a(\boldsymbol{q}_2)e_{\lambda_4}^b(\boldsymbol{k}_2-\boldsymbol{q}_2) \nonumber \\
        &&-q_{2a}q_{2b}e_{\lambda_3}^d(\boldsymbol{q}_2)e_{\lambda_4}^d(\boldsymbol{k}_2-\boldsymbol{q}_2) -2\left(k_{2d}e_{\lambda_3}^d(\boldsymbol{q}_2)\right)q_{2a}e_{\lambda_4}^b(\boldsymbol{k}_2-\boldsymbol{q}_2)\Bigg]\times \left[e_-^\ell(\boldsymbol{k}_1)e_-^m(\boldsymbol{k}_1)e_-^a(\boldsymbol{k}_2)e_-^b(\boldsymbol{k}_2)\right] \nonumber \\
        &&\times\left[e_{-\lambda_1}^{i_1}(\boldsymbol{q}_1)e_{-\lambda_4}^{i_4}(\boldsymbol{k_2}-\boldsymbol{q}_2)P_{i_1i_4}(q_1)\right]\times\left[e_{-\lambda_2}^{i_2}(\boldsymbol{k}_1-\boldsymbol{q}_1)e_{-\lambda_3}^{i_3}(\boldsymbol{q}_2)P_{i_2i_3}(|\boldsymbol{k}_1-\boldsymbol{q}_1|)\right] \nonumber \\
        &&\times |\Pi^{(V)}(q_1)|^2|\Pi^{(V)}(|\boldsymbol{k}_1-\boldsymbol{q}_1|)|^2\times\delta^{(3)}(\boldsymbol{k}_1-\boldsymbol{q}_1+\boldsymbol{q}_2)\times\delta^{(3)}(\boldsymbol{q}_1+\boldsymbol{k}_2-\boldsymbol{q}_2) \\
        && \nonumber \\
        &&= \delta^{(3)}(\boldsymbol{k}_1+\boldsymbol{k}_2)\times\dfrac{8(16\pi G)^4\eta_1^2\eta_2^2}{a(\eta_1)^4a(\eta_2)^4}\sum\limits_{\lambda_1,\lambda_2=\pm}\int\dfrac{d^3q_1}{(2\pi)^{3/2}}\dfrac{|\Pi^{(V)}(q_1)|^2|\Pi^{(V)}(|\boldsymbol{k}_1-\boldsymbol{q}_1|)|^2}{q_1^2|\boldsymbol{k}_1-\boldsymbol{q}_1|^2} \nonumber \\
        &&\times \left[e_-^\ell(\boldsymbol{k}_1)e_-^m(\boldsymbol{k}_1)e_-^a(\boldsymbol{k}_1)^*e_-^b(\boldsymbol{k}_1)^*\right]\times\Bigg[\left(\boldsymbol{q}_1.\left(\boldsymbol{q}_1-\boldsymbol{k}_1\right)+\dfrac{q_1^2|\boldsymbol{q}_1-\boldsymbol{k}_1|^2}{6(1+w)\mathcal{H}_1^2}\right)e_{\lambda_1}^\ell(\boldsymbol{q}_1)e_{\lambda_2}^m(\boldsymbol{k}_1-\boldsymbol{q}_1) \nonumber \\
        &&-q_{1\ell}q_{1m}e_{\lambda_1}^c(\boldsymbol{q}_1)e_{\lambda_2}^c(\boldsymbol{k}_1-\boldsymbol{q}_1) -2\left(k_{1c}e_{\lambda_1}^c(\boldsymbol{q}_1)\right)q_{1\ell}e_{\lambda_2}^m(\boldsymbol{k}_1-\boldsymbol{q}_1)\Bigg]\times\Bigg[\left(\boldsymbol{q}_1.\left(\boldsymbol{q}_1-\boldsymbol{k}_1\right)+\dfrac{q_1^2|\boldsymbol{q}_1-\boldsymbol{k}_1|^2}{6(1+w)\mathcal{H}_2^2}\right) \nonumber \\
        &&\times e_{\lambda_2}^a(\boldsymbol{k}_1-\boldsymbol{q}_1)^*e_{\lambda_1}^b(\boldsymbol{q}_1)^* -q_{1a}q_{1b}e_{\lambda_1}^d(\boldsymbol{q}_1)^*e_{\lambda_2}^d(\boldsymbol{k}_1-\boldsymbol{q}_1)^* +2\left(k_{1d}e_{\lambda_2}^d(\boldsymbol{k}_1-\boldsymbol{q}_1)^*\right)q_{1a}e_{\lambda_1}^b(\boldsymbol{q}_1)^*\Bigg]\:. \nonumber \\
    \end{eqnarray}
Note that the sign of the last term of the second factor term in the product is different from that of the first, which is not an error but results from the consistent use of the $\delta$-functions. Upon expanding the product and simplifying, the expression above becomes
    \begin{eqnarray}
        &&\langle S_+(\boldsymbol{k}_1,\eta_1)S_+(\boldsymbol{k}_2,\eta_2)\rangle_B \nonumber \\
        &&= \delta^{(3)}(\boldsymbol{k}_1+\boldsymbol{k}_2)\times\dfrac{8(16\pi G)^4\eta_1^2\eta_2^2}{a(\eta_1)^4a(\eta_2)^4}\sum\limits_{\lambda_1,\lambda_2=\pm}\int\dfrac{d^3q_1}{(2\pi)^{3/2}}\dfrac{|\Pi^{(V)}(q_1)|^2|\Pi^{(V)}(|\boldsymbol{k}_1-\boldsymbol{q}_1|)|^2}{q_1^2|\boldsymbol{k}_1-\boldsymbol{q}_1|^2} \nonumber \\
        &&\times\Bigg[\left(\boldsymbol{q}_1.\left(\boldsymbol{q}_1-\boldsymbol{k}_1\right)+\dfrac{q_1^2|\boldsymbol{q}_1-\boldsymbol{k}_1|^2}{6(1+w)\mathcal{H}_1^2}\right)\left(\boldsymbol{q}_1.\left(\boldsymbol{q}_1-\boldsymbol{k}_1\right)+\dfrac{q_1^2|\boldsymbol{q}_1-\boldsymbol{k}_1|^2}{6(1+w)\mathcal{H}_2^2}\right) \nonumber \\
        &&\times|e_{\lambda_1}^\ell(\boldsymbol{q}_1)e_-^\ell(\boldsymbol{k}_1)|^2|e_{\lambda_2}^m(\boldsymbol{k}_1-\boldsymbol{q}_1)e_-^m(\boldsymbol{k}_1)|^2+|q_{1\ell}e_-^\ell(\boldsymbol{k}_1)|^4|e_{\lambda_1}^c(\boldsymbol{q}_1)e_{\lambda_2}^c(\boldsymbol{k}_1-\boldsymbol{q}_1)|^2 \nonumber \\
        &&-4|q_{1\ell}e_-^\ell(\boldsymbol{k}_1)|^2k_{1c}e_{\lambda_1}^c(\boldsymbol{q}_1)k_{1d}e_{\lambda_2}^d(\boldsymbol{k}_1-\boldsymbol{q}_1)^*e_-^m(\boldsymbol{k}_1)e_{\lambda_2}^m(\boldsymbol{k}_1-\boldsymbol{q}_1)e_-^b(\boldsymbol{k}_1)^*e_{\lambda_1}^b(\boldsymbol{q}_1)^* \nonumber \\
        &&-2\left(\boldsymbol{q}_1.\left(\boldsymbol{q}_1-\boldsymbol{k}_1\right)+\dfrac{q_1^2|\boldsymbol{q}_1-\boldsymbol{k}_1|^2}{6(1+w)\mathcal{H}_1^2}\right) \nonumber \\
        &&\times\Re\left[\left(q_{1a}e_-^a(\boldsymbol{k}_1)^*\right)^2e_-^\ell(\boldsymbol{k}_1)e_{\lambda_1}^\ell(\boldsymbol{q}_1)e_-^m(\boldsymbol{k}_1)e_{\lambda_2}^m(\boldsymbol{k}_1-\boldsymbol{q}_1)e_{\lambda_1}^d(\boldsymbol{q}_1)^*e_{\lambda_2}^d(\boldsymbol{k}_1-\boldsymbol{q}_1)^*\right] \nonumber \\
        &&+2\left(\boldsymbol{q}_1.\left(\boldsymbol{q}_1-\boldsymbol{k}_1\right)+\dfrac{q_1^2|\boldsymbol{q}_1-\boldsymbol{k}_1|^2}{6(1+w)\mathcal{H}_1^2}\right) \nonumber \\
        &&\times|e_-^b(\boldsymbol{k}_1)e_{\lambda_1}^b(\boldsymbol{q}_1)|^2\left(k_{1d}e_{\lambda_2}^d(\boldsymbol{k}_1-\boldsymbol{q}_1)^*\right)\left(q_{1a}e_-^a(\boldsymbol{k}_1)^*\right)\left(e_-^m(\boldsymbol{k}_1)e_{\lambda_2}^m(\boldsymbol{k}_1-\boldsymbol{q}_1)\right) \nonumber \\
        &&-2\left(\boldsymbol{q}_1.\left(\boldsymbol{q}_1-\boldsymbol{k}_1\right)+\dfrac{q_1^2|\boldsymbol{q}_1-\boldsymbol{k}_1|^2}{6(1+w)\mathcal{H}_2^2}\right) \nonumber \\
        &&\times|e_-^a(\boldsymbol{k}_1)e_{\lambda_2}^a(\boldsymbol{k}_1-\boldsymbol{q}_1)|^2\left(k_{1c}e_{\lambda_1}^c(\boldsymbol{q}_1)\right)\left(q_{1\ell}e_-^\ell(\boldsymbol{k}_1)\right)\left(e_-^b(\boldsymbol{k}_1)^*e_{\lambda_1}^b(\boldsymbol{q}_1)^*\right) \nonumber \\
        &&-2\left(q_{1\ell}e_-^\ell(\boldsymbol{k}_1)\right)^2\left(k_{1d}e_{\lambda_2}^d(\boldsymbol{k}_1-\boldsymbol{q}_1)^*\right)\left(q_{1a}e_-^a(\boldsymbol{k}_1)^*\right)\left(e_-^b(\boldsymbol{k}_1)^*e_{\lambda_1}^b(\boldsymbol{q}_1)^*\right)\left(e_{\lambda_1}^c(\boldsymbol{q}_1)e_{\lambda_2}^c(\boldsymbol{k}_1-\boldsymbol{q}_1)\right) \nonumber \\
        &&+2\left(q_{1a}e_-^a(\boldsymbol{k}_1)^*\right)^2\left(k_{1c}e_{\lambda_1}^c(\boldsymbol{q}_1)\right)\left(q_{1\ell}e_-^\ell(\boldsymbol{k}_1)\right)\left(e_-^m(\boldsymbol{k}_1)e_{\lambda_2}^m(\boldsymbol{k}_1-\boldsymbol{q}_1)\right)\left(e_{\lambda_1}^d(\boldsymbol{q}_1)^*e_{\lambda_2}^d(\boldsymbol{k}_1-\boldsymbol{q}_1)^*\right) \Bigg]\:. \nonumber \\
        &&
    \end{eqnarray}
It is not readily apparent how the last four terms above can be combined and written as the purely real parts of some contractions, unlike in the case of $\langle S_+S_+\rangle_A$. However, this is not an issue, as further simplifications and integrations lead finally to purely real quantities. The correlators corresponding to the three other helicity combinations, \emph{i.e.}, $\langle S_+S_-\rangle$, $\langle S_-S_+\rangle$, and $\langle S_-S_-\rangle$, can be computed in a similar manner, whose explicit forms are not provided here for conciseness. In fact, we find $\langle S_-S_-\rangle=\langle S_+S_+\rangle$ and $\langle S_+S_-\rangle=\langle S_-S_+\rangle$. In the end, the sum of all four helicity combinations, with two Wick terms in each combination, contributes to the total GW energy density spectrum.

In the polar coordinate basis where $\boldsymbol{k_1}$ is aligned along the positive $z$-axis, all these expressions are independent of the azimuthal angle $\phi$, which makes the integral over $\phi$ trivial and equivalent to an overall $2\pi$-scaling. For subsequent numerical calculations of the remaining integrals over $q_1$ and $\theta$, it is convenient to switch to the two dimensionless variables $v=q_1/k_1$ and $u=|\boldsymbol{q_1}-\boldsymbol{k_1}|/k_1$ (which also entails a corresponding Jacobian factor). For a finite integration range $k\in\left[k_{\rm IR},k_{\rm UV}\right]$, the limits on these variables are given by $u\in\left[\max\left(k_{\rm IR}/k_1,|1-v|\right),\min\left(k_{\rm UV}/k_1,1+v\right)\right]$ and $v\in\left[k_{\rm IR}/k_1,k_{\rm UV}/k_1\right]$.

\section{Avoiding magnetic backreaction in the post-inflationary era} \label{sec:appb}

The condition that the magnetic field does not backreact on the background energy density of the Universe at the end of reheating translates to the following inequality \cite{Papanikolaou:2024cwr}:
\begin{equation}
    \int\limits_{k_{\rm reh}}^{k_{\rm inf}}\dfrac{d\rho_B(k,\eta_{\rm reh})}{d\ln k}d\ln k\:<\:\dfrac{\pi^2}{30}g_{\rm reh}T_{\rm reh}^4\:,
\end{equation}
where the magnetic energy density ($\rho_B$) is related to the magnetic power spectrum as
\begin{equation}
    \dfrac{d\rho_B(k,\eta)}{d\ln k}=\dfrac{k^3}{(2\pi)^3}P_B(k,\eta)\:.
\end{equation}
Since the magnetic power spectrum at present time is of the form $P_B(k)=A_Bk^{n_{\rm B}}$ with $A_B$ and $n_{\rm B}$ as given in \eqref{eq:PMF_spectrum_specific}, the spectrum at the end of reheating is obtained via the scaling $P_B(k,\eta_{\rm reh})=P_B(k)/a(\eta_{\rm reh})^4$. One may then plug this expression into the inequality above and calculate the momentum integral analytically to obtain
\begin{equation}
    \dfrac{A_B}{n_{\rm B}+3}\left(k_{\rm inf}^{n_{\rm B}+3}-k_{\rm reh}^{n_{\rm B}+3}\right)\:<\:\dfrac{\pi^2(2\pi)^3}{30}g_{\rm reh}\left(a_{\rm reh}T_{\rm reh}\right)^4\:.
\end{equation}
Throughout the RD era, the temperature scales as $T\propto1/a$ with the scale factor, yielding $a_{\rm reh}T_{\rm reh}=a_{\rm eq}T_{\rm eq}$, where the quantities on the right hand side are evaluated at matter-radiation equality. This enables us to estimate the right hand side of the inequality above, given $g_{\rm reh}\sim100$, $a_{\rm eq}\sim10^{-3}$, and $T_{\rm eq}\sim10^{-9}$ GeV \cite{Tomberg:2021ajh}. Plugging the expressions from \eqref{eq:PMF_spectrum_specific} on the left and using $s=2$, the inequality above then reduces roughly to
\begin{equation}
    \dfrac{1}{2}\left(a_{\rm inf}H_{\rm inf}\right)^2\left(k_{\rm inf}^2-k_{\rm reh}^2\right)\:<\:10^{-42}\:\textrm{GeV}^4\:.
\end{equation}
We may now recall that the scales are defined as $k_{\rm inf}=a_{\rm inf}H_{\rm inf}$ and $k_{\rm reh}=a_{\rm reh}H_{\rm reh}$. Moreover, we have $a_{\rm reh}=a_{\rm inf}e^{\Delta N_{\rm reh}}$ and $H_{\rm reh}=H_{\rm inf}e^{-3\Delta N_{\rm reh}}$ for a kination-dominated reheating epoch. Furthermore, $a_{\rm inf}$ itself is a function of $H_{\rm inf}$ and $\Delta N_\mathrm{reh}$ as given in \eqref{eq:ainf}. Substituting these expressions consecutively in the equality above and simplifying, we arrive at
\begin{equation}
    H_{\rm inf}e^{\Delta N_{\rm reh}}\:<\:10^{51}\:\textrm{GeV}\:,
\end{equation}
where we have assumed $\Delta N_{\rm reh}\gtrsim\mathcal{O}(1)$ to discard some terms. This simple-looking inequality, which connects our two fundamental parameters, must be satisfied to avoid backreaction from the PMF sector at the end of reheating. It is easy to see that even for $H_{\rm inf}\sim10^{15}$ GeV and $\Delta N_{\rm reh}\sim20$, both of which are significantly larger than any of the values considered in our main analysis, the inequality is still comfortably satisfied. Thus, there is no risk of magnetic backreaction in our adopted scenario.

More generally, keeping $s$ free, the backreaction constraint assumes the form
\begin{equation}
    \dfrac{2^{2(s-1)}}{\pi(3-s)}\Gamma\left(s+\dfrac{1}{2}\right)^2\left(a_{\rm inf}H_{\rm inf}\right)^4\left[1-e^{-4(3-s)\Delta N_{\rm reh}}\right]\:<\:10^{-42}\:\textrm{GeV}^4\:,
\end{equation}
which reduces to the case illustrated above for $s=2$, which is relevant to our analysis. However, depending on whether $s<3$ or $s>3$, the constraint may be tighter. For example, for $s=4$, one obtains instead $H_{\rm inf}e^{3\Delta N_{\rm reh}}<10^{50}\:\textrm{GeV}$, which places a significantly lower upper bound on $H_{\rm inf}$ for the same value of $\Delta N_{\rm reh}$, compared to the case for $s=2$. Thus, while the VIGW spectrum may not be overly sensitive to the choice of $s$, we note that backreaction constraints during reheating may be instrumental in ruling out parts of the $\{H_{\rm inf},\Delta N_{\rm reh}\}$ parameter space in a kination-dominated background, particularly for larger values of $s$ which lie beyond the scope of this work. 

\section{Full source term for induced second order tensor perturbations} \label{sec:appc}

In writing \eqref{eq:Source_VV}, we have focused only on the pure vector part of the entire source term for second order tensor perturbations. The full source term, on the other hand, is much more complicated, and contains all possible couplings among first order scalar, vector, and tensor modes. Specifically, there are scalar-scalar, scalar-vector, scalar-tensor, vector-tensor, and tensor-tensor coupling terms, in addition to the vector-vector term which is the focus of this study. While a full analysis considering all the possible terms would be rather tedious and cumbersome, we provide here, for the interested reader, the full second order source that one obtains in the Newtonian gauge via \texttt{xPand} by accounting for all these terms \footnote{Note that, in this section exclusively, we use $h_{ij}$ to denote the first order tensor perturbation, unlike in Sec. \ref{sec:VIGW} where it has been used to denote the second order tensor perturbation.}. 

\begin{align}\label{eq:fullsourceapp}
    S_{ij}=&-V_a\partial_a(\partial_i V_j+\partial_j V_i)+\partial_aV_i\partial_aV_j+\partial_iV_a\partial_jV_a+2V_a\partial_i\partial_jV_a+\dfrac{\Delta V_i\Delta V_j}{6\mathcal{H}^2(1+w)} \\
    & -2\partial_i\phi\partial_j\phi-6\partial_i\psi\partial_j\psi-4\left(\phi\partial_i\partial_j\phi+\psi\partial_i\partial_j\psi\right)+2\left(\partial_i\phi\partial_j\psi+\partial_j\phi\partial_i\psi\right) \nonumber \\
    & +\dfrac{8}{3\mathcal{H}^2(1+w)}\left[\mathcal{H}^2\partial_i\phi\partial_j\phi+\mathcal{H}\left(\partial_i\phi\partial_j\psi'+\partial_j\phi\partial_i\psi'\right)+\partial_i\psi'\partial_j\psi'\right] \\
    & +4h_{ia}'h_{ja}'-2\partial_ih_{ab}\partial_jh_{ab}+4h_{ab}\partial_b\left(\partial_ih_{aj}+\partial_jh_{ai}\right) \nonumber \\
    & +4\partial_bh_{ia}\left(\partial_ah_{bj}-\partial_bh_{aj}\right) -4h_{ab}\left(\partial_i\partial_jh_{ab}+\partial_a\partial_b h_{ij}\right)+12\mathcal{H}^2wh_{ij} \\
    & -\left(\phi'+\psi'+4\mathcal{H}\phi\right)\left(\partial_iV_j+\partial_jV_i\right)-2\phi\left(\partial_iV_j'+\partial_jV_i'\right)+4\mathcal{H}\left(V_i\partial_j\psi+V_j\partial_i\psi\right) \nonumber \\
    & +2\left[\left(V_i\partial_j\psi\right)'+\left(V_j\partial_i\psi\right)'\right]-\dfrac{2}{3\mathcal{H}(1+w)}\left(\Delta V_i\partial_j+\Delta V_j\partial_i\right)\left(\phi+\dfrac{\psi'} {\mathcal{H}}\right) \\
    & +4\phi \left(h_{ij}''+2\mathcal{H} h_{ij}'+6w\mathcal{H}^2h_{ij}\right)-2\left(\phi'-\psi'\right)h_{ij}'-8\mathcal{H}\left(\phi'+3\psi'\right)h_{ij} \nonumber \\
    & -12\psi''h_{ij}+4\left(\psi\Delta h_{ij}-\Delta\phi h_{ij}+2\Delta\psi h_{ij}\right)-4\left(h_{ia}\partial_a\partial_j\psi+h_{ja}\partial_a\partial_i\psi\right) \nonumber \\
    & +2\partial_ah_{ij}\partial_a\left(\phi+3\psi\right)-2\left(\partial_ih_{aj}+\partial_jh_{ia}\right)\partial_a\left(\phi+\psi\right)  \\
    & +4V_a\partial_ah_{ij}'+2\left(V_a'+4\mathcal{H}V_a\right)\partial_ah_{ij}-2\left(h_{ia}'\partial_aV_j+h_{aj}'\partial_aV_i\right) \nonumber \\
    & -2\left[\left(V_a\partial_ih_{aj}\right)'+\left(V_a\partial_jh_{ia}\right)'\right]-4\mathcal{H}V_a\left(\partial_ih_{aj}+\partial_jh_{ia}\right)\:.
\end{align}
In the expression above, we have separately denoted the parts containing first order vector-vector (1st line), scalar-scalar (2nd \& 3rd lines), tensor-tensor (4th \& 5th lines), scalar-vector (6th \& 7th lines), scalar-tensor (8th to 10th lines), and vector-tensor (11th \& 12th lines) couplings. Moreover, while writing this source term, we have ignored terms proportional to $\delta_{ij}$, which are eliminated \emph{\`{a} la} TT-projection.

\bibliographystyle{JHEP}
\bibliography{bibfinal.bib}

\end{document}